\documentclass[12pt,oneside]{amsproc}
\usepackage[english]{babel}
\usepackage{amssymb,epic,eepic}
\renewcommand{\Re}{\operatorname{Re}}
\renewcommand{\Im}{\operatorname{Im}}
\makeatletter
\@addtoreset{equation}{section}
\makeatother

\newtheorem{tm}{Theorem}[section]
\newtheorem{lem}{Lemma}[section]
\title[Spectral portraits]{Spectral portraits of the Orr--Sommerfeld operator
with large Reynolds numbers.}
\author{A.\,A.~Shkalikov}
\thanks{This work is supported by RFFI, the grants no.~01-01-00691 and
no.~01-15-96100.}
\begin{document}
\begin{abstract}
A model problem of the form
\begin{gather*}
	-i\varepsilon y''+q(x)y=\lambda y\\
	y(-1)=y(1)=0
\end{gather*}
is associated with well-known in hydrodynamics Orr--Sommerfeld operator. Here
\(\lambda\) is the spectral parameter, \(\varepsilon\) is the small parameter
which is proportional to the viscocity of the liquid and to the reciprocal of
the Reynolds number, and \(q(x)\) is the velocity of the stationary flow of
the liquid in the channel \(|x|\leqslant 1\). We study the behaviour of the
spectrum of the corresponding model operator as \(\varepsilon\to 0\) with
linear, quadratic and monotonous analytic functions. We show that the sets of
the accumulation points of the spectra (the limit spectral graphs) of the
model and the corresponding Orr--Sommerfeld operators coincide as well as the
main terms of the counting eigenvalue functions along the curves of the
graphs.
\end{abstract}
\maketitle
The well-known in hydrodynamics Orr--Sommerfeld equation is obtained by
linearization of the Navier--Stokes equation in the infinite three-dimension
spatial layer \((x,\xi,\eta)\in\mathbb R^3\), where \(|x|\leqslant 1\) and
\((\xi,\eta)\in\mathbb R^2\), assuming that a stationary unperturbed solution
for the velocity profile is of the form \((q(x),0,0)\). This equation with
respect to a function \(y=y(x)\) has the form (see details in the book of
Drazin and Reid~\cite{DR}, for example)
\begin{equation}\label{eq1}
	(D^2-\alpha^2)^2y-i\alpha R\left[q(x)(D^2-\alpha^2)-q''(x)\right]y=
	-i\alpha R\lambda(D^2-\alpha^2)y.
\end{equation}
Here \(D=d/dx\), \(\alpha\) is a wave number (\(\alpha\neq 0\)) appearing
after the separation of the variables \((\xi,\eta)\in\mathbb R^2\), \(R\)
is the Reynolds number characterizing the viscosity of the liquid and
\(\lambda\) is the spectral parameter. Usually the boundary conditions
\begin{equation}\label{eq2}
	y(\pm 1)=y'(\pm 1)=0
\end{equation}
are associated with equation~\eqref{eq1}.

The main goal of this paper is to describe qualitatively the limit behaviour
of the spectrum of problem~\eqref{eq1},~\eqref{eq2} as \(R\to\infty\). The
Reynolds number \(R\) is proportional to the reciprocal of the viscosity.
Therefore, our problem is devoted to the description of the spectrum of the
Orr--Sommerfeld problem for the liquid which is almost ideal. During long time
it was supposed that the spectrum of the Rayleigh problem
\begin{equation}\label{eq:0.3}
\begin{gathered}
	q(x)(D^2-\alpha^2)y-q''(x)y=\lambda(D^2-\alpha^2)y,\\
	y(-1)=y(1)=0,
\end{gathered}
\end{equation}
plays an important role for the solution of the problem which is raised
above. The Rayleigh problem is obtained (after dividing of equation~\eqref{eq1}
by \(-i\alpha R\)) by formal passing to the limit as \(R\to\infty\) and by
eliminating of the "superfluous" boundary conditions. A vast literature is
devoted to the study of the spectrum of the problem~\eqref{eq:0.3}. The papers
of Lin~\cite{L} and the cited book~\cite{DR} can be recommended as a guide in
this subject. Actually, the spectrum of the Rayleigh problem plays no
essential role in the description of the spectral portraits of the
Orr--Sommerfeld problem as \(R\to\infty\). This will be seen in the sequel.

It is known~\cite{DR}, that the spectrum of the problem~\eqref{eq:0.3} consists
of the segment \([m,M]\), where \(m\) and \(M\) are minimum and maximum of the
function \(q(x)\) (it is supposed that \(q(x)\) is continuous), and possibly,
some isolated eigenvalues of finite multiplicity lying outside this segment.
Perhaps, Heisenberg was the first who noticed that there is no continuity
between the spectra of the Orr--Sommerfeld problem with large \(R\) and the
spectrum of the Rayleigh problem. It may happen that there is a domain
containing the interval \((m,M)\) which is free of the spectra of
problem~\eqref{eq1},~\eqref{eq2} for all sufficiently large \(R\). This
phenomenon was named as a "Heisenberg tongue". In~1924 Heisenberg proved the
existence of a system of fundamental solutions for equation~\eqref{eq1} having
special representation (see~\cite{DR}). This result is very essential for
explanation of such a phenomenon. However, we are not aware of Heisenberg'
papers containing ideas to explain the phenomenon. It was Morawetz~\cite{M},
who proved that in the case of the Couette profile \(q(x)=x\) the spectrum of
the problem~\eqref{eq1},~\eqref{eq2} is concentrated in a
\mbox{\(\delta\)-neigh}\-bour\-hoods of the ray \([-i/\sqrt3,-i\infty)\), of
two segments \([\pm 1,-i/\sqrt3]\) and the isolated eigenvalues \(\{\mu_k\}\)
of the Rayleigh problem for \(q(x)=x\). Actually, the Rayleigh problem for
\(q(x)=x\) has no isolated eigenvalues (see~\cite{DR}), therefore, the last
reservation can be droped. Moreover, Morawetz made an attempt to prove a
similar result for more general functions. She assumed (this assumption is
quite essential in her method) that \(q(x)\) is an entire function with real
values on the real axis and maps bijectively the whole complex plane \(\mathbb
C\) into itself. Probably, it was not noticed that all the function possessing
these properties have the representation \(q(x)=\Theta x+\theta\), where
\(0\neq\Theta\), \(\theta\in\mathbb R\). There is another important problem
which was left beyond the paper~\cite{M}. What is the set of the accumulation
points of the eigenvalues as \(R\to\infty\)? Morawetz emphasized
in~\cite{M} that her method did not allow to get the information if the
eigenvalues exists as \(R\to\infty\) near each point of the segments \([\pm
1,-i/\sqrt3]\).

In 90-ies there appeared papers (see~\cite{Tr}, \cite{RSH}, for example), where
a simplier problem of the form
\begin{gather}\label{eq3}
	-i\varepsilon z''+q(x)z=\lambda z,\\ \label{eq4}
	z(-1)=z(1)=0,
\end{gather}
was associated with~\eqref{eq1},~\eqref{eq2}. Here \(\varepsilon\) and
\(\lambda\) are the small and spectral parameters, respectively. One can
consider this problem as a simplified model for~\eqref{eq1},~\eqref{eq2}. The
following arguments can confirm this point.

Let us make in equation~\eqref{eq1} the substitution \(z=(D^2-\alpha^2)y\).
From this equation taking into account the conditions \(y(-1)=y'(-1)=0\), we
find 
\begin{equation}\label{eq5}
	y(x)=\dfrac{1}{\alpha^2}\int\limits_{-1}^{x}\sinh\alpha(x-\xi)
	z(\xi)\,d\xi.
\end{equation}
Then equation~\eqref{eq1} takes the form
\begin{equation}\label{eq6}
	-i\varepsilon(D^2-\alpha^2)z+q(x)z+Kz=\lambda z,
\end{equation}
where
\begin{align*}
	Kz&=\int\limits_{-1}^{x}\sinh 2(x-\xi)q''(\xi)z(\xi)\,d\xi,&
	\varepsilon&=\dfrac{1}{\alpha R}.
\end{align*}
It follows from~\eqref{eq5} that \(y(-1)=y'(-1)=0\). Therefore, one can rewrite
boundary conditions~\eqref{eq2} in the form
\begin{align}\label{eq7}
	\int\limits_{-1}^{1}z(\xi)\sinh\alpha(1-\xi)\,d\xi&=0,&
	\int\limits_{-1}^{1}z(\xi)\cosh\alpha(1-\xi)\,d\xi&=0.
\end{align}
Hence, problem~\eqref{eq1},~\eqref{eq2} is equivalent to
problem~\eqref{eq6},~\eqref{eq7}. This reduction was realized by Orr in~1915.
Now, if we neglect the influence of the integral operator \(K\) (notice, that
\(K=0\) in the case \(q(x)=x\)) and assume that boundary conditions do
not change essentially the spectral portraits as \(\varepsilon\to 0\), we come
to model problem~\eqref{eq3},~\eqref{eq4} up to the shift of the spectral
parameter by \(i\varepsilon\alpha^2\). Of course, these arguments are
heuristic. However, we shall show below that the similarity of the spectral
portraits of the model and the Orr--Sommerfeld problem as \(\varepsilon\to 0\)
can be proved rigorously.

If we put \(\varepsilon>0\) in equation~\eqref{eq3} instead of
\(i\varepsilon\), we get a selfadjoint problem with small parameter. Such a
problem is well investigated long ago (see~\cite{RSS}, for exemple). Its
spectrum is real, it condenses as \(\varepsilon\to 0\) and the formulae for
the eigenvalue localization can be written out explicitly. They are known as
the \textit{Bohr--Sommerfeld quantization formulae}.

The replacement of \(\varepsilon\) by \(i\varepsilon\) changes the problem
dramatically. In~1997 the author~\cite{Sh1} described the spectral portraits of
model problem~\eqref{eq3},~\eqref{eq4} as \(\varepsilon\to 0\) in the case
\(q(x)=x\) and paid attention to the fact that for analytic functions \(q(x)\)
the spectrum is concentrated along some curves which are determined by the
geometry of the Stokes lines of equation~\eqref{eq3}. It is useful here to
recall the definitions. The zeros of the equation \(q(z)-\lambda=0\) in the
complex \mbox{\(z\)-pla}\-ne are called \textit{turning points} and the lines
\[
	\gamma_{\xi_{\lambda}}=\left\{z\in\mathbb C\;\vline\;
	\Re\int_{\xi_{\lambda}}^{z}\sqrt{i(q(\xi)-\lambda)}\,d\xi=0\right\},
\]
outgoing from a fixed turning point \(\xi_{\lambda}\) are called \textit{the
Stokes lines} of equation~\eqref{eq3}. These lines come either to the boundary
of a domain \(G\), where \(q(z)\) is holomorphic (in particular, go to
\(\infty\) if \(q(z)\) is an entire function), or end in other turning points.
The maximal connected set consisting of the Stokes lines and containing a
turning point \(\xi_{\lambda}\) is called \textit{the Stokes complex}
corresponding to \(\xi_{\lambda}\). The Stokes complex may contain several
turning points but not necessary all of them. The union of Stokes complexes
corresponding to all turning points is called \textit{the Stokes graph}.

As we have mentioned before, for analytic (or piece-wise analytic) functions
\(q(x)\) the spectrum of problem~\eqref{eq3},~\eqref{eq4} is concentrated
near some curves in the \mbox{\(\lambda\)-pla}\-ne as \(\varepsilon\to 0\).
We call them \textit{limit spectral curves}. The union of the limit spectral
curves we call \textit{limit spectral graph}. One should not mix the limit
spectral curves or the limit spectral graph with the Stokes lines and the
Stokes graph which lie in \mbox{\(z\)-pla}\-ne. Concrete forms of the limit
spectral curves can be found comparatively easy only for the linear function
\(q(x)=x\). For nonlinear functions \(q(x)\) these curves take a complicated
form. Therefore, attempting to describe the spectral portraits for
non-selfadjoint problems with small or large parameter one meets serious
difficulties even in the case of model problem~\eqref{eq3},~\eqref{eq4}. Now
we can formulate our goals more concretely.
\begin{enumerate}\itshape
\item To find functions \(q(x)\) of particular and general form for which the
spectral portraits of problem~\eqref{eq3},~\eqref{eq4} as \(\varepsilon\to 0\)
can be described completely. Of course, special attention should be paid to
profiles which correspond to stationary solutions of the Navier--Stokes
equation, in particular to the Couette profile \(q(x)=x\), Poiseuille profile
\(q(x)=1-x^2\) and Couette--Poiseuille profile \(q(x)=ax^2+bx+c\), 
\(a,b,c\in\mathbb R\);
\item To find the formulae for the eigenvalues near the limit spectral curves
as \(\varepsilon\to 0\), provided that these curves are already determined;
\item To solve the same problems for the original Orr--Sommerfeld problem with
the same functions \(q(x)\).
\end{enumerate}

At the present moment the problems, which we have raised up, are solved only
partially. In this paper we present the obtained results based on the research
of the author and his students A.~V.~Dyachenko, S.~N.~Tumanov and
M.~I.Neiman-zade. We should also mention recent papers of Chapman~\cite{Ch},
Redparth~\cite{Re} and Stepin~\cite{St} which are closely connected with the
problems in question.

\section{Model problem: the case \(q(x)=x\).}\label{par1}
Let us consider the spectral problem
\begin{gather}\label{eq1:1}
	-i\varepsilon y''=(x-\lambda)y,\\ \label{eq1:2}
	y(-1)=y(1)=0,
\end{gather}
where \(\varepsilon\) is a small parameter. As we have mentioned, computer 
calculations give a striking result: the eigenvalues of this problem for
sufficiently small \(\varepsilon\to 0\) are concentrated on the ray
\(\gamma_{\infty}=[-i/\sqrt3,-i\infty)\) and near the segments
\(\gamma_{\pm}=[\pm 1,-i/\sqrt3]\) (see~Figure~\ref{fig1}). Of course, the
density of the eigenvalues is increased as \(\varepsilon\to 0\). Based on the
method of the paper of Morawetz~\cite{M} (although
problem~\eqref{eq1:1},~\eqref{eq1:2} was not treated in~\cite{M}) one can
show, that given \(\delta>0\) the spectrum of this problem  is contained in
the \mbox{\(\delta\)-neigh}\-bour\-hood of the limit spectral graph
\[
	\Gamma=\gamma_{+}\cup\gamma_{-}\cup\gamma_{\infty},
\]
provided that \(\varepsilon<\varepsilon_0(\delta)\).
The author was not acquainted with the paper~\cite{M} when he started to study
this subject. In~\cite{Sh1} he described the spectral portrait of 
problem~\eqref{eq1},~\eqref{eq2} as \(\varepsilon\to 0\) based on the
properties of fundamental solution of the Airy equation. Moreover, the
explicit formulae for the eigenvalues near \(\gamma_+\) and \(\gamma_-\) were
written down. Later on, these results were completed and sharpened in the
papers of Djachenko and Shkalikov~\cite{DSh1},~\cite{DSh2}. Here we will
formulate the last results concerning problem~\eqref{eq1:1},~\eqref{eq1:2}
and sketch basic ideas.

\begin{figure}[t]
\setlength{\unitlength}{0.00027489in}
\begingroup\makeatletter\ifx\SetFigFont\undefined%
\gdef\SetFigFont#1#2#3#4#5{%
  \reset@font\fontsize{#1}{#2pt}%
  \fontfamily{#3}\fontseries{#4}\fontshape{#5}%
  \selectfont}%
\fi\endgroup%
{\renewcommand{\dashlinestretch}{30}
\begin{picture}(9024,10389)(0,-10)
\put(6762,7932){\blacken\ellipse{90}{90}}
\put(6762,7932){\ellipse{90}{90}}
\put(8112,8697){\blacken\ellipse{90}{90}}
\put(8112,8697){\ellipse{90}{90}}
\put(7707,8472){\blacken\ellipse{90}{90}}
\put(7707,8472){\ellipse{90}{90}}
\put(7347,8247){\blacken\ellipse{90}{90}}
\put(7347,8247){\ellipse{90}{90}}
\put(7077,8112){\blacken\ellipse{90}{90}}
\put(7077,8112){\ellipse{90}{90}}
\put(6402,7752){\blacken\ellipse{90}{90}}
\put(6402,7752){\ellipse{90}{90}}
\put(6177,7617){\blacken\ellipse{90}{90}}
\put(6177,7617){\ellipse{90}{90}}
\put(5952,7482){\blacken\ellipse{90}{90}}
\put(5952,7482){\ellipse{90}{90}}
\put(5682,7347){\blacken\ellipse{90}{90}}
\put(5682,7347){\ellipse{90}{90}}
\put(5457,7212){\blacken\ellipse{90}{90}}
\put(5457,7212){\ellipse{90}{90}}
\put(5232,7077){\blacken\ellipse{90}{90}}
\put(5232,7077){\ellipse{90}{90}}
\put(5052,6942){\blacken\ellipse{90}{90}}
\put(5052,6942){\ellipse{90}{90}}
\put(4827,6852){\blacken\ellipse{90}{90}}
\put(4827,6852){\ellipse{90}{90}}
\put(4647,6717){\blacken\ellipse{90}{90}}
\put(4647,6717){\ellipse{90}{90}}
\put(2217,7932){\blacken\ellipse{90}{90}}
\put(2217,7932){\ellipse{90}{90}}
\put(867,8697){\blacken\ellipse{90}{90}}
\put(867,8697){\ellipse{90}{90}}
\put(1272,8472){\blacken\ellipse{90}{90}}
\put(1272,8472){\ellipse{90}{90}}
\put(1632,8247){\blacken\ellipse{90}{90}}
\put(1632,8247){\ellipse{90}{90}}
\put(1902,8112){\blacken\ellipse{90}{90}}
\put(1902,8112){\ellipse{90}{90}}
\put(2577,7752){\blacken\ellipse{90}{90}}
\put(2577,7752){\ellipse{90}{90}}
\put(2802,7617){\blacken\ellipse{90}{90}}
\put(2802,7617){\ellipse{90}{90}}
\put(3027,7482){\blacken\ellipse{90}{90}}
\put(3027,7482){\ellipse{90}{90}}
\put(3297,7347){\blacken\ellipse{90}{90}}
\put(3297,7347){\ellipse{90}{90}}
\put(3522,7212){\blacken\ellipse{90}{90}}
\put(3522,7212){\ellipse{90}{90}}
\put(3747,7077){\blacken\ellipse{90}{90}}
\put(3747,7077){\ellipse{90}{90}}
\put(3927,6942){\blacken\ellipse{90}{90}}
\put(3927,6942){\ellipse{90}{90}}
\put(4152,6852){\blacken\ellipse{90}{90}}
\put(4152,6852){\ellipse{90}{90}}
\put(4332,6717){\blacken\ellipse{90}{90}}
\put(4332,6717){\ellipse{90}{90}}
\put(4512,6447){\blacken\ellipse{90}{90}}
\put(4512,6447){\ellipse{90}{90}}
\put(4512,6222){\blacken\ellipse{90}{90}}
\put(4512,6222){\ellipse{90}{90}}
\put(4512,5997){\blacken\ellipse{90}{90}}
\put(4512,5997){\ellipse{90}{90}}
\put(4512,5772){\blacken\ellipse{90}{90}}
\put(4512,5772){\ellipse{90}{90}}
\put(4512,5547){\blacken\ellipse{90}{90}}
\put(4512,5547){\ellipse{90}{90}}
\put(4512,5322){\blacken\ellipse{90}{90}}
\put(4512,5322){\ellipse{90}{90}}
\put(4512,5097){\blacken\ellipse{90}{90}}
\put(4512,5097){\ellipse{90}{90}}
\put(4512,4872){\blacken\ellipse{90}{90}}
\put(4512,4872){\ellipse{90}{90}}
\put(4512,4602){\blacken\ellipse{90}{90}}
\put(4512,4602){\ellipse{90}{90}}
\put(4512,4377){\blacken\ellipse{90}{90}}
\put(4512,4377){\ellipse{90}{90}}
\put(4512,4152){\blacken\ellipse{90}{90}}
\put(4512,4152){\ellipse{90}{90}}
\put(4512,3927){\blacken\ellipse{90}{90}}
\put(4512,3927){\ellipse{90}{90}}
\put(4512,3702){\blacken\ellipse{90}{90}}
\put(4512,3702){\ellipse{90}{90}}
\put(4512,3477){\blacken\ellipse{90}{90}}
\put(4512,3477){\ellipse{90}{90}}
\put(4512,3252){\blacken\ellipse{90}{90}}
\put(4512,3252){\ellipse{90}{90}}
\put(4512,3027){\blacken\ellipse{90}{90}}
\put(4512,3027){\ellipse{90}{90}}
\put(4512,2802){\blacken\ellipse{90}{90}}
\put(4512,2802){\ellipse{90}{90}}
\put(4512,2577){\blacken\ellipse{90}{90}}
\put(4512,2577){\ellipse{90}{90}}
\put(4512,2352){\blacken\ellipse{90}{90}}
\put(4512,2352){\ellipse{90}{90}}
\put(4512,2127){\blacken\ellipse{90}{90}}
\put(4512,2127){\ellipse{90}{90}}
\put(4512,1902){\blacken\ellipse{90}{90}}
\put(4512,1902){\ellipse{90}{90}}
\put(4512,1677){\blacken\ellipse{90}{90}}
\put(4512,1677){\ellipse{90}{90}}
\put(4512,1452){\blacken\ellipse{90}{90}}
\put(4512,1452){\ellipse{90}{90}}
\put(4512,1227){\blacken\ellipse{90}{90}}
\put(4512,1227){\ellipse{90}{90}}
\put(4512,9012){\ellipse{180}{180}}
\put(8562,9012){\ellipse{180}{180}}
\put(462,9012){\ellipse{180}{180}}
\put(4512,6672){\blacken\ellipse{90}{90}}
\put(4512,6672){\ellipse{90}{90}}
\put(4512,1047){\blacken\ellipse{90}{90}}
\put(4512,1047){\ellipse{90}{90}}
\put(4512,822){\blacken\ellipse{90}{90}}
\put(4512,822){\ellipse{90}{90}}
\put(4512,597){\blacken\ellipse{90}{90}}
\put(4512,597){\ellipse{90}{90}}
\put(4512,372){\blacken\ellipse{90}{90}}
\put(4512,372){\ellipse{90}{90}}
\put(4512,147){\blacken\ellipse{90}{90}}
\put(4512,147){\ellipse{90}{90}}
\path(12,9012)(9012,9012)
\path(4512,10362)(4512,12)
\put(4737,9282){\makebox(0,0)[lb]{\(0\)}}
\put(8832,9237){\makebox(0,0)[lb]{\(1\)}}
\put(282,9372){\makebox(0,0)[lb]{\(-1\)}}
\put(4962,6267){\makebox(0,0)[lb]{\(-i/\sqrt3\)}}
\end{picture}
}
\caption{}\label{fig1}
\end{figure}

First, let us note about the following simple fact.
\begin{lem}\label{lem1:-1}
For any \(\varepsilon>0\) the spectrum of problem~\eqref{eq1:1},~\eqref{eq1:2}
lies in the closure of the semi-strip
\[
	\Pi=\left\{\lambda\;\vline\;\Im\lambda<0,\,-1<\Re\lambda<1\right\}.
\]
\end{lem}
\begin{proof}
Let us consider the operator \(Ly=i\varepsilon y''+xy\) generated by boundary
conditions~\eqref{eq1:2}. Its spectrum coincide with the spectrum of the
problem. Obviously, the values of the quadratic form
\[
	(Ly,y)=-i\varepsilon(y',y')+(xy,y)
\]
lie in \(\Pi\), provided that \(\|y\|=1\) and \(y\) belongs to the domain
\(\mathfrak D(L)\) of the operator. The values of the quadratic form form
the numerical range of the operator \(L\). Now, the assertion follows from the
well-known fact: the spectrum of any operator lies in the closure of its
numerical range.
\end{proof}

The substitution \(\xi=(-i\varepsilon)^{-1/3}(x-\lambda)\) reduces
problem~\eqref{eq1:1},~\eqref{eq1:2} to the following one:
\begin{gather*}
	z''(\xi)=\xi z(\xi),\\
	z(\xi_1)=z(\xi_2)=0,\\
	\xi_1=e^{i\pi/6}\varepsilon^{-1/3}(-1-\lambda),\quad
	\xi_2=e^{i\pi/6}\varepsilon^{-1/3}(1-\lambda).
\end{gather*}
Thus, we have the spectral problem for the Airy equation. This problem is
uncommon, since the spectral parameter is involved only in the boundary
conditions, and the end points change depending on the spectral parameter.
However, the eigenvalues of this problem can be found by standard means.
Namely, the eigenvalues are determined by the equation
\[
	\Delta(\lambda)=\begin{vmatrix}v(\xi_1)&v(\xi_2)\\
	w(\xi_1)&w(\xi_2)\end{vmatrix}=0,
\]
where \(v(\xi)\) and \(w(\xi)\) are linear independent solutions of the Airy
equation. Let us take the Airy function \(v(\xi)\) subject the following
asymptotic equality
\begin{equation}\label{eq1:3}
	v(\xi)=\dfrac{1}{2\sqrt{\pi}\xi^{1/4}}e^{-\frac23\xi^{3/2}}
	\left(1+O\left(|\xi|^{-3/2}\right)\right),\quad\xi\in
	\Lambda_{\pi-\delta},\quad\xi\to\infty,
\end{equation}
where \(\Lambda_{\pi-\delta}=\left\{\xi\;\vline\;|\arg\xi|<\pi-\delta
\right\}\), \(\delta>0\), and the main branches of \(\xi^{1/4}\) and
\(\xi^{3/2}\) are chosen in \(\Lambda_{\pi-\delta}\). A domain, where
representation~\eqref{eq1:3} is valid, can be enlarged. We take advantage from
the following result obtained in~\cite{DSh1}.
\begin{lem}\label{lem1:1}
The Airy function \(v(\xi)\) preserves asymptotic representation~\eqref{eq1:3}
in the domain
\[
	\Lambda=\left\{\xi\;\vline\;|\xi|>1,\;|\arg\xi|<\pi-\dfrac34
	|\xi|^{-3/2}\ln|\xi|\right\},
\]
i.~e.~\eqref{eq1:3} is valid for all \(\xi\in\Lambda\), and the remainder is
majorated by \(C|\xi|^{-3/2}\) uniformly for \(\xi\in\Lambda\).
\end{lem}
\begin{proof}
We make use of the well-known representation~\cite{O}
\begin{equation}\label{eq1:4}
	v(\xi)=e^{-\pi i/3}v\left(e^{2\pi i/3}\xi\right)+e^{\pi i/3}
	v\left(e^{-2\pi i/3}\xi\right).
\end{equation}
Notice, that in the intersection of \(\Lambda\) with the second quadrant of the
complex plane the estimate
\[
	\left|v\left(e^{-2\pi i/3}\xi\right)\right|\leqslant
	C|\xi|^{-3/2}\left|v\left(e^{2\pi i/3}\xi\right)\right|
\]
is valid. Moreover, the first summand in~\eqref{eq1:4} has
representation~\eqref{eq1:3} in this intersection. Using the symmetry principle
we obtain the assertion of Lemma.
\end{proof}

Let us consider the function
\begin{equation}\label{eq1:5}
	f(\lambda)=\int\limits_{-1}^1 e^{-\pi i/4}\sqrt{x-\lambda}\,dx=
	\dfrac23 e^{-\pi i/4}\cdot\left[(1-\lambda)^{3/2}-
	(-1-\lambda)^{3/2}\right],
\end{equation}
which will be used in the formulation of the main theorem of this section. The
function \(f(\lambda)\) is holomorphic in the domain \(\Pi\) where the spectrum
is located. We choose the branch by the condition \(f(-i/\sqrt3)>0\).

Denote
\[
	\Lambda_{\alpha}=\left\{\lambda\in\mathbb C\;\vline\;
	|\arg\lambda|<\alpha\right\}.
\]
Fix an arbitrary number \(\sigma>2^{-2/3}3^{-1/4}\) and consider the domain
\(D_{\sigma}\) (see~Figure~\ref{fig2}) bounded by the lines \(\Re\lambda=\pm
1\) from the right and from the left, and by the lines passing through the
points \(\pm 1\) and the point
\begin{equation}\label{eq1:6}
	d_{\sigma}=-i\left(\dfrac{1}{\sqrt3}+\sigma
	\varepsilon^{1/2}|\ln\varepsilon)|\right)
\end{equation}
from the above. The constant \(2^{-2/3}3^{-1/4}\) for the choice of \(\sigma\)
is explicit in our calculations, we can not make it smaller.

\begin{figure}[t]
\setlength{\unitlength}{0.00027489in}
\begingroup\makeatletter\ifx\SetFigFont\undefined%
\gdef\SetFigFont#1#2#3#4#5{%
  \reset@font\fontsize{#1}{#2pt}%
  \fontfamily{#3}\fontseries{#4}\fontshape{#5}%
  \selectfont}%
\fi\endgroup%
{\renewcommand{\dashlinestretch}{30}
\begin{picture}(9024,10399)(0,-10)
\thicklines
\put(4512,6682){\ellipse{720}{720}}
\thinlines
\put(6762,7942){\blacken\ellipse{90}{90}}
\put(6762,7942){\ellipse{90}{90}}
\put(8112,8707){\blacken\ellipse{90}{90}}
\put(8112,8707){\ellipse{90}{90}}
\put(7707,8482){\blacken\ellipse{90}{90}}
\put(7707,8482){\ellipse{90}{90}}
\put(7347,8257){\blacken\ellipse{90}{90}}
\put(7347,8257){\ellipse{90}{90}}
\put(7077,8122){\blacken\ellipse{90}{90}}
\put(7077,8122){\ellipse{90}{90}}
\put(6402,7762){\blacken\ellipse{90}{90}}
\put(6402,7762){\ellipse{90}{90}}
\put(6177,7627){\blacken\ellipse{90}{90}}
\put(6177,7627){\ellipse{90}{90}}
\put(5952,7492){\blacken\ellipse{90}{90}}
\put(5952,7492){\ellipse{90}{90}}
\put(5682,7357){\blacken\ellipse{90}{90}}
\put(5682,7357){\ellipse{90}{90}}
\put(5457,7222){\blacken\ellipse{90}{90}}
\put(5457,7222){\ellipse{90}{90}}
\put(5232,7087){\blacken\ellipse{90}{90}}
\put(5232,7087){\ellipse{90}{90}}
\put(5052,6952){\blacken\ellipse{90}{90}}
\put(5052,6952){\ellipse{90}{90}}
\put(4827,6862){\blacken\ellipse{90}{90}}
\put(4827,6862){\ellipse{90}{90}}
\put(4647,6727){\blacken\ellipse{90}{90}}
\put(4647,6727){\ellipse{90}{90}}
\put(2217,7942){\blacken\ellipse{90}{90}}
\put(2217,7942){\ellipse{90}{90}}
\put(867,8707){\blacken\ellipse{90}{90}}
\put(867,8707){\ellipse{90}{90}}
\put(1272,8482){\blacken\ellipse{90}{90}}
\put(1272,8482){\ellipse{90}{90}}
\put(1632,8257){\blacken\ellipse{90}{90}}
\put(1632,8257){\ellipse{90}{90}}
\put(1902,8122){\blacken\ellipse{90}{90}}
\put(1902,8122){\ellipse{90}{90}}
\put(2577,7762){\blacken\ellipse{90}{90}}
\put(2577,7762){\ellipse{90}{90}}
\put(2802,7627){\blacken\ellipse{90}{90}}
\put(2802,7627){\ellipse{90}{90}}
\put(3027,7492){\blacken\ellipse{90}{90}}
\put(3027,7492){\ellipse{90}{90}}
\put(3297,7357){\blacken\ellipse{90}{90}}
\put(3297,7357){\ellipse{90}{90}}
\put(3522,7222){\blacken\ellipse{90}{90}}
\put(3522,7222){\ellipse{90}{90}}
\put(3747,7087){\blacken\ellipse{90}{90}}
\put(3747,7087){\ellipse{90}{90}}
\put(3927,6952){\blacken\ellipse{90}{90}}
\put(3927,6952){\ellipse{90}{90}}
\put(4152,6862){\blacken\ellipse{90}{90}}
\put(4152,6862){\ellipse{90}{90}}
\put(4332,6727){\blacken\ellipse{90}{90}}
\put(4332,6727){\ellipse{90}{90}}
\put(4512,6457){\blacken\ellipse{90}{90}}
\put(4512,6457){\ellipse{90}{90}}
\put(4512,6232){\blacken\ellipse{90}{90}}
\put(4512,6232){\ellipse{90}{90}}
\put(4512,6007){\blacken\ellipse{90}{90}}
\put(4512,6007){\ellipse{90}{90}}
\put(4512,5782){\blacken\ellipse{90}{90}}
\put(4512,5782){\ellipse{90}{90}}
\put(4512,5557){\blacken\ellipse{90}{90}}
\put(4512,5557){\ellipse{90}{90}}
\put(4512,5332){\blacken\ellipse{90}{90}}
\put(4512,5332){\ellipse{90}{90}}
\put(4512,5107){\blacken\ellipse{90}{90}}
\put(4512,5107){\ellipse{90}{90}}
\put(4512,4882){\blacken\ellipse{90}{90}}
\put(4512,4882){\ellipse{90}{90}}
\put(4512,4612){\blacken\ellipse{90}{90}}
\put(4512,4612){\ellipse{90}{90}}
\put(4512,4387){\blacken\ellipse{90}{90}}
\put(4512,4387){\ellipse{90}{90}}
\put(4512,4162){\blacken\ellipse{90}{90}}
\put(4512,4162){\ellipse{90}{90}}
\put(4512,3937){\blacken\ellipse{90}{90}}
\put(4512,3937){\ellipse{90}{90}}
\put(4512,3712){\blacken\ellipse{90}{90}}
\put(4512,3712){\ellipse{90}{90}}
\put(4512,3487){\blacken\ellipse{90}{90}}
\put(4512,3487){\ellipse{90}{90}}
\put(4512,3262){\blacken\ellipse{90}{90}}
\put(4512,3262){\ellipse{90}{90}}
\put(4512,3037){\blacken\ellipse{90}{90}}
\put(4512,3037){\ellipse{90}{90}}
\put(4512,2812){\blacken\ellipse{90}{90}}
\put(4512,2812){\ellipse{90}{90}}
\put(4512,2587){\blacken\ellipse{90}{90}}
\put(4512,2587){\ellipse{90}{90}}
\put(4512,2362){\blacken\ellipse{90}{90}}
\put(4512,2362){\ellipse{90}{90}}
\put(4512,2137){\blacken\ellipse{90}{90}}
\put(4512,2137){\ellipse{90}{90}}
\put(4512,1912){\blacken\ellipse{90}{90}}
\put(4512,1912){\ellipse{90}{90}}
\put(4512,1687){\blacken\ellipse{90}{90}}
\put(4512,1687){\ellipse{90}{90}}
\put(4512,1462){\blacken\ellipse{90}{90}}
\put(4512,1462){\ellipse{90}{90}}
\put(4512,1237){\blacken\ellipse{90}{90}}
\put(4512,1237){\ellipse{90}{90}}
\put(4512,9022){\ellipse{180}{180}}
\put(8562,9022){\ellipse{180}{180}}
\put(462,9022){\ellipse{180}{180}}
\put(4512,6682){\blacken\ellipse{90}{90}}
\put(4512,6682){\ellipse{90}{90}}
\put(4512,1057){\blacken\ellipse{90}{90}}
\put(4512,1057){\ellipse{90}{90}}
\put(4512,832){\blacken\ellipse{90}{90}}
\put(4512,832){\ellipse{90}{90}}
\put(4512,607){\blacken\ellipse{90}{90}}
\put(4512,607){\ellipse{90}{90}}
\put(4512,382){\blacken\ellipse{90}{90}}
\put(4512,382){\ellipse{90}{90}}
\put(4512,157){\blacken\ellipse{90}{90}}
\put(4512,157){\ellipse{90}{90}}
\path(12,9022)(9012,9022)
\path(4512,10372)(4512,22)
\thicklines
\path(462,22)(462,3442)(4512,6322)
	(8562,3442)(8562,22)
\put(4737,9292){\makebox(0,0)[lb]{\(0\)}}
\put(8832,9247){\makebox(0,0)[lb]{\(1\)}}
\put(282,9382){\makebox(0,0)[lb]{\(-1\)}}
\put(1777,3702){\makebox(0,0)[lb]{\(D_{\sigma}\)}}
\end{picture}
}
\caption{}\label{fig2}
\end{figure}

\begin{lem}\label{lem1:2}
The function \(f(\lambda)\) has the following properties.
\begin{enumerate}
\item \(f(\lambda)\) is holomorphic in the lower half-plane and takes real
values on the negative imaginary axis;
\item for all \(\lambda\in D_{\sigma}\) the values of \(f(\lambda)\) and
\(-if'(\lambda)\) belong to the sector \(\Lambda_{\pi/6}\), and the values of
\(f''(\lambda)\) belong to the lower half-plane;
\item \(f(\lambda)\) increases monotonously and tends to \(\infty\) as
\(\lambda\to -i\infty\) along the negative imaginary axis;
\item \(f(\lambda)=2\sqrt{i\lambda}+O\left(|\lambda|^{-3/2}\right)\),
\(f'(\lambda)=i/\sqrt{i\lambda}+O\left(|\lambda|^{-3/2}\right)\) as
\(\lambda\to\infty\), \(\lambda\in D_{\sigma}\);
\item \(f(D_{\sigma})<\Re f(\lambda)\) for \(\lambda\in
D_{\sigma}\);
\item \(|f(\lambda_1)-f(\lambda_2)|\geqslant
c\frac{|\lambda_1-\lambda_2|}{\sqrt{|\lambda_1|+|\lambda_2|}}\) for
\(\lambda_1,\lambda_2\in D_{\sigma}\), where the constant \(c\) does not
depend on \(\lambda_1\), \(\lambda_2\);
\item the function \(f(\lambda)\) takes real values in \(D_{\sigma}\) only at
the imaginary axis.
\end{enumerate}
\end{lem}
\begin{proof}
All these assertions follow easily from representation~\eqref{eq1:5}. The
details can be found in~\cite{DSh2}.
\end{proof}

Let \(\{r_k\}_{k=1}^{\infty}\) be the increasing sequence of the zeros of the
function \(v(-r)\), where \(v(\xi)\) is the Airy function with
representation~\eqref{eq1:3}. Then \(r_k>0\), and the asymptotic equalities
\begin{equation}\label{eq:rk}
	r_k=\left(\dfrac{3\pi}{2}\left(k-\dfrac14\right)\right)^{2/3}+
	O\left(\dfrac{1}{k^{4/3}}\right),\quad k=1,2,\ldots
\end{equation}
are valid. It is not obviuos, that the numeration in these formulas
starts from \(k=1\). The explanations are given
in~\cite{DSh2}.

Set
\begin{align*}
	\delta_{\sigma}&=\delta_{\sigma}(\varepsilon)=\sigma\varepsilon
	|\ln\varepsilon|,&\sigma&>2^{-3/2}3^{-1/4}.
\end{align*}
Denote
\begin{equation}\label{eq:mu}
\begin{aligned}
	\mu_k^-&=-1+e^{-\pi i/6}\varepsilon^{1/3}r_k,&
	k&=1,2,\ldots,k_1,k_1+1\\
	\mu_k^+&=1-e^{\pi i/6}\varepsilon^{1/3}r_k,&
	k&=1,2,\ldots,k_1,k_1+1,
\end{aligned}
\end{equation}
where the number \(k_1=k_1(\varepsilon)\) is the largest integer such that the
inequality \(\varepsilon^{1/3}r_{k_1}<2/\sqrt3-\delta_{\sigma}\) holds. The
numbers \(\mu_k^-\) and \(\mu_k^+\) lie on the segments
\(\gamma_-=[1,-i/\sqrt3]\) and \(\gamma_+=[-1,-i/\sqrt3]\) respectively. They
come up to the \mbox{\(\delta_{\sigma}\)-neigh}\-bour\-hood of the knot-point
\(-i/\sqrt3\), but only the last ones get inside this neighbourhood.

Next, let us construct a sequence of numbers on the imaginary axis. By virtue
of Lemma~\ref{lem1:2} \(f(-i\rho)\) is the increasing function for
\(\rho\in\mathbb R^+\). Moreover,
\[
	f(d_{\sigma})=f(-i(1/\sqrt3+\delta_{\sigma}))>0.
\]
Given \(\varepsilon>0\) choose the smallest integer \(k_0=k_0(\varepsilon)\)
such that \(f(d_{\sigma})<\pi k_0\varepsilon^{1/2}\). Then, each equation
\begin{align*}
	f(-i\rho)-\pi k\varepsilon^{1/2}&=0,& k&=k_0-1,k_0,k_0+1,\ldots,
\end{align*}
has the only solution \(\rho=\rho_k>-id_{\sigma}\) for \(k\geqslant k_0\)
and \(\rho_{k_0-1}<-id_{\sigma}\). The sequence
\(\{\rho_k\}_{k_0-1}^{\infty}\) is monotonous and \(\rho_k\to\infty\) as
\(k\to\infty\).

Now, we are ready to formulate the main result on the eigenvalue portrait of
model problem~\eqref{eq1:1},~\eqref{eq1:2}.
\begin{tm}\label{tm1:1}
Let \(\sigma>2^{-3/2}3^{-3/4}\) be any fixed number and
\(\delta_{\sigma}=\sigma\varepsilon|\ln\varepsilon|\). Set
\begin{align*}
	\varphi(t)&=\dfrac43\Re\left(2e^{\pi i/6}-t\right)^{3/2},&
	t&\in[0,2/\sqrt3].
\end{align*}
Denote by \(U_k^\pm\) the neighbourhoods of the points \(\mu_k^{\pm}\),
\(1\leqslant k\leqslant k_1\), of radius \(\gamma_k=C\exp\left(
-\varepsilon^{-1/2}\varphi\left(\varepsilon^{1/2}r_k\right)\right)\), by
\(U_k^{\infty}\) the neighbourhoods of the points \(-i\rho_k\), \(k\geqslant
k_0-1\), of radius \(C\rho_k^{-1}\varepsilon\), and by \(U_0\) the
\mbox{\(\delta_{\sigma}\)-neigh}\-bour\-hood of the knot-point \(-i/\sqrt3\).

There are numbers \(C=C(\sigma)\) and \(\varepsilon_0=\varepsilon_0(\sigma)\)
not depending on \(\varepsilon\) and \(k\) such that for all
\(\varepsilon<\varepsilon_0\) the circles \(\{U_k^{\pm}\}_1^{k_1+1}\) and
\(\{U_k^{\infty}\}_{k_0-1}^{\infty}\) contain the only eigenvalue of
problem~\eqref{eq1:1},~\eqref{eq1:2}. Moreover, the eigenvalues in
\(U_k^{\infty}\) are pure imaginary. All the other eigenvalues lie in the
circle \(U_0\).

In other words, the pure imaginary eigenvalues in the domain \(D_{\sigma}\)
have asymptotics
\begin{align}\label{eq:im}
	\lambda_k&=-i\left(\rho_k+\varepsilon\rho_k^{-1}O(1)\right),&
	k&=k_0-1,k_0,k_0+1,\ldots,
\end{align}
the eigenvalues near the segments \(\gamma_{\pm}\) have asymptotics
\begin{align}\label{eq:<}
	\lambda_k^{\pm}&=\mu_k^{\pm}+\exp\left(-\varepsilon^{-1/2}
	\varphi\left(\varepsilon^{1/2}r_k\right)\right)O(1),&
	k&=1,2,\ldots,k_1,k_1+1,
\end{align}
and there are
\[
	\sigma 2^{1/2}3^{3/4}|\ln\varepsilon|+O(1)
\]
eigenvalues inside the circle \(U_0\). In all these formulae the quantity
\(|O(1)|\) is estimated by a constant \(C\) depending only on \(\sigma\).
\end{tm}
\begin{proof}
Here we outline only the main ideas, the details can be found in~\cite{DSh2}.

Let us construct the characteristic determinant of the problem using the Airy
solutions \(v(\xi)\) and \(w(\xi)=v\left(e^{-2\pi i/3}\xi\right)\). We have 
\[
	\Delta(\xi)=v(\xi_1)v\left(e^{-2\pi i/3}\xi_2\right)-
	v\left(e^{-2\pi i/3}\xi_1\right)v(\xi_2).
\]

\textit{Step~1.} First, we investigate the zeros of \(\Delta(\lambda)\) in the
domain \(D_{\sigma}\). Notice that the variables \(\xi_j\) and \(e^{-2\pi
i/3}\xi_j\), \(j=1,2\), lie in the domain \(\Lambda\) of Lemma~\ref{lem1:2},
provided that \(\lambda\in D_{\sigma}\). Therefore, for all \(\lambda\in
D_{\sigma}\), we can use asymptotic representation~\eqref{eq1:3} for the
functions \(v(\xi_j)\) and \(v\left(e^{-2\pi i/3}\xi_j\right)\). After simple
transformations we get the equation
\[
	e^{4/3\left(\xi_2^{3/2}-\xi_1^{3/2}\right)}=e^{2i\varepsilon^{-1/2}
	f(\lambda)}=1+\lambda^{-3/2}\varepsilon^{1/2}O(1),
\]
which is equivalent in the domain \(D_{\sigma}\) to the equation
\(\Delta(\lambda)=0\). Taking the logarithm of both sides, we come to the
equations
\begin{align}\label{eq:f}
	f(\lambda)-\pi k\varepsilon^{1/2}&=\lambda^{-3/2}
	\varepsilon^{1/2}O(1),& k&\in\mathbb Z,&
	\lambda&\in D_{\sigma}.
\end{align}
Now, recall the definition of the number \(k_0\) and the estimate
\(|f'(\lambda)|>C|\lambda|^{-1/2}\), \(\lambda\in D_{\sigma}\), which follows
from Lemma~\ref{lem1:2}. Then, using the Rouch\'e theorem, we find that the
equation~\eqref{eq:f} has the only root \(\lambda_k\) inside the circle
\(U_k^{\infty}\), provided that \(k\geqslant k_0+1\). This root is necessarily
pure imaginary, since the spectrum of the problem is symmetric with respect to
the imaginary axis.

Actually, we can prove the same for the indices \(k_0\) and \(k_0-1\) if we
consider a larger domain \(D_{\sigma'}\) with
\(2^{-2/3}3^{-1/4}<\sigma'<\sigma\). It is easily seen due to properties of
\(f(\lambda)\) proved in Lemma~\ref{lem1:2}, that there are no roots in the
domain \(D_{\sigma}\) outside the circles
\(\{U_k^{\infty}\}_{k_0-1}^{\infty}\).

\textit{Step~2.} While investigating the zeros of \(\Delta(\lambda)\) near the
segment \([-1,-i/\sqrt3]\) it is convenient to rewrite the equation
\(\Delta(\lambda)=0\) in the form
\begin{equation}\label{eq:b.e}
	\dfrac{v(\xi_1)}{v\left(e^{-2\pi i/3}\xi_1\right)}=
	\dfrac{v(\xi_2)}{v\left(e^{-2\pi i/3}\xi_2\right)}.
\end{equation}
Let us consider the trapezium \(\Omega_{\sigma}\subset\Pi\), which is bounded
by the lines \(\Re\lambda=-1\), \(\Re\lambda=0\), the real axis and the line
passing through the points \(1\) and
\(\beta_{\sigma}=-1+e^{-\pi/6}(2/\sqrt3-\delta_{\sigma})\). The point
\(\beta_{\sigma}\) is the intersection of the segment \(\gamma_-\) and the
circumference of radius \(\delta_{\sigma}\) centered at the knot-point
\(-i/\sqrt3\). The variables \(\xi_1\), \(\xi_2\), \(e^{-2\pi i/3}\xi_1\),
\(e^{-2\pi i/3}\xi_2\) take the values in the domain \(\Lambda\) of
Lemma~\ref{lem1:1} if \(\lambda\in\Omega_{\sigma}\). Hence, we can use
asymptotics~\eqref{eq1:3} for the functions involved in~\eqref{eq:b.e}. It is
easy to show that the right hand side of~\eqref{eq:b.e} is majorated by
\(C\left|\exp\left(-\frac43\xi_2^{3/2}\right)\right|\). Introducing the new
variable \(r=(\lambda+1)e^{i\pi/6}\) we get from~\eqref{eq:b.e} the relation
\begin{align}\label{eq:v}
	V(r)&:=\dfrac{v\left(-\varepsilon^{-1/3}r\right)}{
	v\left(-e^{-2\pi i/3}r\right)}=O(1)\exp\left(-\varepsilon^{-1/2}
	\varphi(r)\right),&r&=(\lambda+1)e^{i\pi/6}.
\end{align}
The function \(V(r)\) has the simple zeros \(\varepsilon^{1/3}r_k\),
\(k=1,2,\ldots\), where \(r_k\) have asymptotics~\eqref{eq:rk}. Using the
Rouch\'e
theorem (here technicalities are omitted), we obtain that
equation~\eqref{eq:v} has simple roots in the neighbourhoods of the points
\(\varepsilon^{1/3}r_k\) of radius \(\gamma_k=C\exp\left(-\varepsilon^{-1/2}
\varphi\left(\varepsilon^{1/2}r_k\right)\right)\), provided that \(1\leqslant
k\leqslant k_1+1\). Coming back to the variable \(\lambda\), we obtain the
assertion of the theorem about the eigenvalues near the segment \(\gamma_-\).
It is easily seen from~\eqref{eq:v} that there are no other zeros in
\(\Omega_{\sigma}\) except the zeros lying in the circles
\(\{U_k^-\}_1^{k_1+1}\). We pay attention that \(\lambda_k^-\) lie in
exponentially small neighbourhoods of the points \(\mu_k^-\) provided that
\(|\mu_k^--i/\sqrt3|>c>0\). However, the radii \(\beta_k\) of the circles
\(U_k^-\) increase as \(\mu_k^-\to-i/\sqrt3\), since \(\varphi(t)\to 0\) as
\(t\to 2/\sqrt3\). It is easy to calculate that
\(\beta_k<C\varepsilon^{\alpha}\) with some \(\alpha>1/2\) (depending on
\(\sigma\)) for all \(k\leqslant k_1+1\).

\textit{Step~3.} We have to show that there are no eigenvalues in the domain
\(\Pi^-\setminus(\Omega_{\sigma}\cup D_{\sigma})\), where \(\Pi^-\) is the
intersection of the semistrip \(\Pi\) and the left half-plane. Using
representation~\eqref{eq1:3} it is easy to show that for \(\lambda\) belonging
to this domain the inequality
\[
	\left|v(\xi_1)v\left(e^{-2\pi i/3}\xi_2\right)\right|>
	\left|v\left(e^{-2\pi i/3}\xi_1\right)v(\xi_2)\right|
\]
holds.

\textit{Step~4.} Denote by \(N(\rho,\varepsilon)\) the number of the
eigenvalues of the problem~\eqref{eq1:1},~\eqref{eq1:2} lying above the line
\(\Im\lambda=-\rho\). Suppose that \(\rho>1/\sqrt3+\delta\), \(\delta>0\). Fix
numbers \(\varepsilon_0\) and \(\varepsilon\) (\(\varepsilon<\varepsilon_0\))
and define numbers \(k_0\) and \(k\) as the largest integers satisfying the
conditions
\begin{align*}
	\pi\varepsilon_0^{1/2}k_0&<f(-i\rho),&
	\pi\varepsilon^{1/2}k&<f(-i\rho).
\end{align*}
Then
\[
	|N(\rho,\varepsilon)-N(\rho,\varepsilon_0)-(k-k_0)|\leqslant 2.
\]
Therefore,
\[
	N(\rho,\varepsilon)=k+O(1)=\dfrac{1}{\pi\varepsilon^{1/2}}
	f(-i\rho)+O(1).
\]
It is easily seen that \(N(\rho,\varepsilon_0)=k_0=0\) for sufficiently large
\(\varepsilon_0\) and \(O(1)\) in the last formula takes the values \(0\) or
\(\pm 1\).

Now, using asymptotics~\eqref{eq:<} we can calculate the number of the
eigenvalues in the circle \(U_0\):
\[
	N_{\sigma}:=\dfrac{1}{\pi\varepsilon^{1/2}}\left(f\left(
	-\dfrac{i}{\sqrt3}-i\delta_{\sigma}\right)-\dfrac43\left(
	\dfrac{2}{\sqrt3}-\delta_{\sigma}\right)^{3/2}\right)+O(1),
\]
where \(|O(1)|\leqslant 3\). After simple transformations we obtain
\[
	N_{\sigma}=\dfrac{2^{1/2}3^{3/4}\sigma}{\pi}|\ln\varepsilon|+
	O(1)+o(1),
\]
where \(o(1)\to 0\) as \(\varepsilon\to 0\) and \(|O(1)|\leqslant 3\). This
completes the proof.
\end{proof}

In addition to this theorem we shall say some words about the movement of the
\(\lambda_k(\varepsilon)\) as \(\varepsilon\to 0\). Since the function
\(f(-i\rho)\) is monotonous at the semiaxis \(\mathbb R^+\), it follows from
the proof of Theorem~\ref{tm1:1} that all the eigenvalues
\(\lambda_k(\varepsilon)\) lying on the imaginary axis are simple and move up
as \(\varepsilon\to 0\). Due to the symmetry they can jump off the imaginary
axis only in the case when \(\lambda_k\) gets into the circle \(U_0\) and
catches up the preceding eigenvalue \(\lambda_{k-1}\). After the collision
they leave the imaginary axis. We have no information about their behaviour
in the circle \(U_0\). However, if we proceed to diminish \(\varepsilon\), the
eigenvalues go out from the small circle \(U_0\) and move along the segments
\(\gamma_-\) and \(\gamma_+\) coming to these segments exponentially close.

\section{Model problem: the case of a monotonous analytic function \(q(x)\).}
Here we shall show that the spectral portrait of the explicitly solvable model
problem~\eqref{eq1:1},~\eqref{eq1:2} is not accidental; similar phenomena are
observed for a wide class of monotonous analytic functions \(q(x)\).
Certainly, another method will be used for the study of general problem.
Namely, we shall make use of the phase integral method or the so-called
\mbox{WKBJ-meth}\-od. The results of this section were announced by the
author~\cite{Sh2}.

We consider the problem
\begin{gather}\label{eq2:1}
	i\varepsilon^2y''+q(x)y=\lambda y,\\ \label{eq2:2}
	y(-1)=y(1)=0.
\end{gather}
We pay attention that here for convenience the small parameter is denoted by
\(\varepsilon^2\) instead of \(\varepsilon\) in equation~\eqref{eq1:1}. It
might be assumed for the simplicity that \(q(x)\) admits an analytic
continuation in the whole complex plane, although only the analyticity in a
neighbourhood of the segment \([-1,1]\) is required in the sequel in addition
to the conditions formulated below.

Let \([a,b]\) be the range of the function \(q(x)\) defined for \(x\in[-1,1]\). 
Denote by \(L(\varepsilon)\) the operator corresponding to
problem~\eqref{eq2:1},~\eqref{eq2:2}. Obviously, the values of the quadratic
form \((L(\varepsilon)y,y)\) for \(y\in\mathfrak D(L)\), \(\|y\|=1\), lie in
the semistrip
\[
	\Pi=\left\{\lambda\;\vline\;\Im\lambda<0,\;a<\Re\lambda<b\right\}.
\]
Hence, for any \(\varepsilon>0\) the eigenvalues of
problem~\eqref{eq1:1},~\eqref{eq1:2} lie in this semistrip
(see~Lemma~\ref{lem1:1}).

Let us formulate the main assumptions for \(q(x)\).
\begin{itemize}\itshape
\item[(i)] The function \(q(x)\) is real for \(x\in[-1,1]\) and there is a
domain \(G\subset\mathbb C\) such that \(q(z)\) is analytic in \(G\) and maps
\(\overline{G}\) bijectively onto \(\overline{\Pi}\) (here the overline
implies the closure of the domains).
\item[(ii)] For any \(c\in(a,b)\) the preimage of the ray \(r_c=\{\lambda\;
\vline\;\lambda=c-it,\;0\leqslant t<\infty\}\) is a function with respect to
the imaginary axis, i.~e. any line \(\Im\lambda=\mathrm{const}\) either
intersects the preimage of the ray \(r_c\) only once, or has no intersection
points.
\end{itemize}

Condition~(i) implies that \(q(x)\) is strictly monotonous on the
segment \([-1,1]\). Without loss of generality we assume that \(q(x)\) is
increasing. In this case the domain \(G\) belongs entirely to the lower
half-plane (otherwise, we get a contradiction with condition~(ii)). By
virtue of the symmetry principle, the function \(q(z)\) maps bijectively
\(G\cup G^*\cup(-1,1)\) onto the strip \(a<\Re\lambda<b\), where \(G^*\) is
symmetrical to \(G\) with respect to the real axis. In particular, \(q'(x)>0\)
for all \(x\in(-1,1)\). The above conditions are satisfied, for example, for
the functions \(q(x)=\sin(\pi x/2)\), \(q(x)=(x+1)^2\) etc.

Let us consider the following functions in the closure of the domain \(\Pi\):
\begin{align*}
	Q(\lambda)&=\int\limits_{-1}^1\sqrt{i(q(x)-\lambda)}\,dx,&
	\lambda&\in\overline{\Pi},\\
	Q^{\pm}(\lambda)&=\pm
	\int\limits_{\xi_{\lambda}}^{\pm 1}
	\sqrt{i(q(\xi)-\lambda)}\,d\xi,& \lambda&\in\overline{\Pi},
\end{align*}
where \(\xi_{\lambda}\) is the only root of the equation \(q(\xi)-\lambda=0\)
lying in the domain \(G\). The branches of these functions can be chosen
arbitrary. To be certain, fix the branches by the conditions
\begin{gather}\label{eq:Q}
	Q(a)=\int\limits_{-1}^1\sqrt{i(q(x)-a)}\,dx=e^{i\pi/4}\alpha,\qquad
	\alpha>0,\\
	\label{eq:Qpm}
	Q^+(a)=Q(a),\qquad Q^+(\lambda)+Q^-(\lambda)=Q(\lambda).
\end{gather}
Certainly, the functions \(Q\), \(Q^+\) and \(Q^-\) are analytic in \(\Pi\) and
continuous in \(\overline{\Pi}\).

Define the following curves in the semistrip \(\overline{\Pi}\):
\begin{align*}
	\tilde\gamma_{\infty}&=\{\lambda\in\overline{\Pi}
	\;\vline\;\Re Q(\lambda)=0\},\\
	\tilde\gamma_{\pm}&=\{\lambda\in\overline{\Pi}
	\;\vline\;\Re Q^{\pm}(\lambda)=0\}.
\end{align*}
These curves are depicted in Figure~\ref{fig3} for the case
\(q(x)=(x+1)^2/4\).
Some parts of these curves are drawn by dotted lines. We shall show below that
these parts play no role in the description of the spectral portrait. The
remaining parts form \textit{the limit spectral graph}, and each point of
this graph accumulates the eigenvalues of the problem as \(\varepsilon\to 0\).

\begin{figure}[t]
\setlength{\unitlength}{0.00027489in}
\begingroup\makeatletter\ifx\SetFigFont\undefined%
\gdef\SetFigFont#1#2#3#4#5{%
  \reset@font\fontsize{#1}{#2pt}%
  \fontfamily{#3}\fontseries{#4}\fontshape{#5}%
  \selectfont}%
\fi\endgroup%
{\renewcommand{\dashlinestretch}{30}
\begin{picture}(5874,9532)(0,-10)
\put(552,9022){\ellipse{180}{180}}
\put(5052,9022){\ellipse{180}{180}}
\path(552,9022)(5052,9022)(5052,22)
\path(552,22)(552,9022)
\path(12,9022)(5862,9022)
\thicklines
\path(552,8977)(553,8976)(556,8972)
	(562,8967)(571,8958)(583,8945)
	(599,8928)(620,8906)(644,8881)
	(672,8852)(703,8820)(738,8784)
	(774,8746)(812,8707)(851,8666)
	(891,8625)(930,8584)(969,8544)
	(1007,8505)(1045,8466)(1080,8429)
	(1115,8394)(1148,8360)(1179,8328)
	(1209,8298)(1237,8269)(1264,8242)
	(1289,8216)(1313,8192)(1337,8168)
	(1359,8146)(1380,8125)(1401,8104)
	(1421,8085)(1440,8066)(1460,8047)
	(1485,8022)(1511,7998)(1536,7974)
	(1560,7951)(1585,7928)(1610,7905)
	(1636,7882)(1662,7858)(1689,7834)
	(1717,7810)(1746,7784)(1776,7758)
	(1807,7732)(1838,7705)(1869,7678)
	(1899,7653)(1928,7629)(1954,7606)
	(1978,7587)(1997,7570)(2013,7557)
	(2024,7548)(2031,7542)(2035,7539)(2037,7537)
\path(5052,9022)(5050,9021)(5047,9019)
	(5040,9015)(5030,9009)(5016,9001)
	(4996,8989)(4971,8975)(4942,8958)
	(4907,8938)(4867,8915)(4823,8889)
	(4775,8862)(4724,8832)(4670,8801)
	(4615,8769)(4558,8737)(4501,8704)
	(4443,8671)(4387,8639)(4331,8607)
	(4276,8576)(4222,8546)(4171,8517)
	(4121,8489)(4072,8462)(4026,8436)
	(3981,8411)(3937,8387)(3896,8364)
	(3855,8343)(3816,8322)(3778,8301)
	(3741,8282)(3705,8263)(3669,8245)
	(3634,8227)(3599,8209)(3564,8192)
	(3530,8174)(3495,8158)(3460,8141)
	(3426,8124)(3391,8107)(3355,8091)
	(3319,8074)(3282,8057)(3245,8040)
	(3206,8023)(3166,8005)(3125,7987)
	(3083,7969)(3039,7950)(2993,7930)
	(2946,7910)(2897,7889)(2846,7868)
	(2794,7846)(2740,7824)(2685,7801)
	(2630,7778)(2574,7755)(2518,7732)
	(2463,7709)(2409,7687)(2356,7666)
	(2306,7646)(2260,7627)(2217,7610)
	(2178,7594)(2144,7580)(2115,7569)
	(2091,7559)(2072,7551)(2058,7546)
	(2048,7541)(2042,7539)(2038,7538)(2037,7537)
\thinlines
\dashline{150.000}(2037,7537)
	(1586,7350)(1142,7179)
	(654,7027)(552,6997)
\dashline{150.000}(2037,7537)
	(2447,7336)(3672,6781)
	(4453,6477)(5097,6232)
\put(507,9382){\makebox(0,0)[lb]{\(0\)}}
\put(5232,9337){\makebox(0,0)[lb]{\(1\)}}
\put(3657,7762){\makebox(0,0)[lb]{\(\gamma_+\)}}
\put(772,7622){\makebox(0,0)[lb]{\(\gamma_-\)}}
\put(2487,4972){\makebox(0,0)[lb]{\(\gamma_{\infty}\)}}
\thicklines
\path(2037,7537)(2037,7536)(2037,7533)
	(2037,7528)(2038,7520)(2038,7509)
	(2039,7493)(2040,7473)(2041,7448)
	(2042,7418)(2044,7383)(2045,7342)
	(2047,7296)(2049,7244)(2052,7188)
	(2054,7127)(2057,7061)(2060,6992)
	(2063,6919)(2066,6843)(2069,6764)
	(2072,6683)(2075,6601)(2079,6517)
	(2082,6433)(2085,6349)(2089,6264)
	(2092,6180)(2095,6097)(2098,6015)
	(2101,5934)(2104,5854)(2107,5775)
	(2110,5698)(2113,5622)(2115,5548)
	(2118,5476)(2120,5405)(2122,5335)
	(2125,5267)(2127,5200)(2129,5135)
	(2131,5070)(2133,5006)(2134,4944)
	(2136,4882)(2138,4820)(2139,4759)
	(2141,4699)(2142,4638)(2143,4578)
	(2145,4518)(2146,4457)(2147,4397)
	(2148,4336)(2150,4274)(2150,4221)
	(2151,4166)(2152,4111)(2153,4056)
	(2154,4000)(2155,3943)(2156,3886)
	(2156,3828)(2157,3768)(2158,3708)
	(2158,3646)(2159,3583)(2160,3519)
	(2160,3453)(2161,3386)(2161,3317)
	(2162,3246)(2162,3173)(2163,3098)
	(2163,3022)(2164,2943)(2164,2862)
	(2165,2779)(2165,2694)(2166,2606)
	(2166,2517)(2166,2425)(2167,2332)
	(2167,2237)(2167,2139)(2168,2041)
	(2168,1941)(2168,1839)(2169,1737)
	(2169,1634)(2169,1532)(2170,1429)
	(2170,1327)(2170,1226)(2170,1127)
	(2170,1029)(2171,934)(2171,842)
	(2171,754)(2171,669)(2171,589)
	(2171,514)(2171,444)(2172,379)
	(2172,320)(2172,266)(2172,219)
	(2172,177)(2172,141)(2172,111)
	(2172,85)(2172,65)(2172,50)
	(2172,38)(2172,31)(2172,26)
	(2172,23)(2172,22)
\thinlines
\dashline{150.000}(2037,7537)
	(2010,7992)(1984,8429)
	(1950,8966)(1947,9022)
\end{picture}
}
\caption{}\label{fig3}
\end{figure}

First, we prove some important properties of the curves \(\tilde\gamma_{\pm}\)
and \(\tilde\gamma_{\infty}\).

\begin{lem}\label{lem2:1}
The curve \(\tilde\gamma_+\) (\(\tilde\gamma_-\)) passes through the point
\(b\) (point \(a\)) of the semistrip \(\overline{\Pi}\). Both curves are the
functions with respect to the real axis.
\end{lem}
\begin{proof}
Consider the curve \(\tilde\gamma_+\), the proof for \(\tilde\gamma_-\) is
similar. Fix a number \(c\in(a,b)\). By assumption~(i) there is the
point \(\xi_c\in(-1,1)\) such that \(q(\xi_c)-c=0\). Let \(\lambda=c-it\),
\(t\geqslant 0\) and \(\xi_{\lambda}\) be the root of the equation
\(q(\xi)-\lambda=0\), \(\xi_{\lambda}\in G\). We have
\[
	Q^+(\lambda)=\int\limits_{\xi_{\lambda}}^1\sqrt{i(q(\xi)-\lambda)}\,
	d\xi=\left(\int\limits_{\xi_{\lambda}}^{\xi_c}+\int\limits_{\xi_c}^1
	\right)\sqrt{i(q(\xi)-\lambda)}\,d\xi=:F_1(\lambda)+F_2(\lambda).
\]
The branch of the function \(Q^+(\lambda)\) is defined by
condition~\eqref{eq:Qpm}. Since \(Q^+(\lambda)\) is continuous in
\(\overline{\Pi}\), we have
\[
	Q^+(c)=\int\limits_{\xi_{c}}^1\sqrt{i(q(\xi)-c)}\,d\xi=
	e^{\pi i/4}\alpha_c,\quad\alpha_c>0.
\]
Notice, that
\[
	\Re F'_2(\lambda)=-\dfrac12\int\limits_{\xi_c}^1
	\Re\dfrac{e^{\pi i/4}}{\sqrt{(q(\xi)-c)+it}}\,d\xi<0,\quad
	\lambda=c-it,
\]
since \(0<\arg\left[(q(\xi)-c)+it\right]<\pi/2\). Hence, the function \(\Re
F_2(c-it)\) decreases monotonously, moreover, \(\Re F_2(c-it)\to 0\) as
\(t\to\infty\). Further, denote by \(\xi(\mu)\) the preimage of the segment
\(\lambda=c-i\mu\), \(0\leqslant\mu\leqslant t\), under the map \(q(z)\). Due
to assumption~(ii) we can parameterize this curve as follows
\(\xi(s)=-is+r(s)\), \(0\leqslant s=s(\mu)\leqslant s(t)\), \(s'(\mu)
\geqslant 0\).

While calculating the function \(F_1(\lambda)\) we can take the integral along
the curve \(\xi(\mu)\). Then,
\[
	F_1(\lambda)=\int\limits_{\xi_{\lambda}}^{\xi_c}\sqrt{i(q(\xi)-
	\lambda)}\,d\xi=\int\limits_{s(t)}^0\sqrt{i(c-i\mu-\lambda)}\,d\xi(s)=
	\int\limits_t^0\sqrt{\mu-t}\,d(-is(\mu)+r(s(\mu))).
\]
Bearing in mind the choice of the branch, we find that the function
\[
	\Re F_1(c-it)=-\int\limits_0^t\sqrt{t-\mu}\,ds(\mu)
\]
decreases monotonously as \(t\to+\infty\). Therefore, the function \(\Re
Q^+(c-it)\) decreases monotonously as \(t\) changes from \(0\) to \(+\infty\). 
Moreover, this function is positive at zero and negative in a neighbourhood of
\(+\infty\). Hence, the equation \(\Re Q^+(c-it)=0\) has the only root
\(t_c<0\), and the curve \(\tilde\gamma_+\) is a function with respect to the
real axis. Since \(Q^+(\lambda)\to 0\) as \(\lambda\to b\), this curve ends in
the point \(b\). Lemma is proved.
\end{proof}
\begin{lem}\label{lem2:2}
The curve \(\tilde\gamma_{\infty}\) is a function with respect to the imaginary
axis.
\end{lem}
\begin{proof}
The branch of \(Q(\lambda)\) in \(\overline{\Pi}\) is defined by
condition~\eqref{eq:Q}. Since \(a<q(x)<b\) for \(x\in(-1,1)\), we find that
for all \(t>0\)
\begin{align*}
	\Re Q(a-it)&=\Re\int\limits_{-1}^1\sqrt{i(q(x)-a+it)}\,dx>0,\\
	\Re Q(b-it)&=\Re\int\limits_{-1}^1\sqrt{i(q(x)-b+it)}\,dx<0.
\end{align*}
Further,
\begin{equation}\label{eq2:3}
	\Re Q'(\lambda)=-\dfrac12\int_{-1}^1\Re\dfrac{e^{\pi i/4}}{\sqrt{q(x)-
	\lambda}}\,dx<0,
\end{equation}
since \(\Im(q(x)-\lambda)>0\) for \(x\in(-1,1)\) and \(\lambda\in\Pi\). This
implies that for any fixed \(t>0\) the function \(Q(c-it)\) of the variable
\(c\in[-1,1]\) vanishes in the only point. Lemma is proved.
\end{proof}
\begin{lem}\label{lem2:3}
The functions \(Q^+(\lambda)\), \(Q^-(\lambda)\), \(Q(\lambda)\) are univalent
in the semistrip \(\Pi\). In particular, the functions \(\Im Q^+(\lambda)\),
\(\Im Q^-(\lambda)\), \(\Im Q(\lambda)\) are strictly monotonous along the
curves \(\tilde\gamma_+\), \(\tilde\gamma_-\) and
\(\tilde\gamma_{\infty}\),
respectively.
\end{lem}
\begin{proof}
It was shown in Lemma~\ref{lem2:1} that \(\Re\frac{d}{dt}Q^+(c-it)<0\) for
\(t>0\) and any fixed \(c\in(a,b)\). Consequently, \(\Re\frac{d}{d\lambda}
Q^+(\lambda)>0\) for \(\lambda\in\Pi\). This implies the univalence of
\(Q^+(\lambda)\) in \(\Pi\). Since \(\Re Q^+(\lambda)=0\) along the curve
\(\tilde\gamma_+\), the derivative of the function
\(\Im Q^+(\lambda)\) along this
curve does not vanish and preserves the sign. The same assertion is true for
the function \(Q^-(\lambda)\). The univalence of the function \(Q(\lambda)\)
in \(\Pi\) follows from inequality~\eqref{eq2:3}. Then, the function \(\Im
Q(\lambda)\) is strictly monotonous along the curve \(\tilde\gamma_{\infty}\).
Lemma is proved.
\end{proof}
\begin{lem}\label{lem2:4}
\(\Re Q^+(\lambda)>0\) if \(\lambda\) lies above the curve \(\tilde\gamma_+\)
and \(\Re Q^+(\lambda)<0\) if \(\lambda\) lies below this curve. Similarly,
\(\Re Q^-(\lambda)<0\) (\(>0\)) if \(\lambda\) lies above (below) the curve
\(\tilde\gamma_-\).
\end{lem}
\begin{proof}
This assertion follows from the proof of
Lemma~\ref{lem2:1} and representation~\eqref{eq:Qpm}.
\end{proof}

\begin{lem}\label{lem2:6}
The curves \(\tilde\gamma_+\) and \(\tilde\gamma_-\) have the inly intersection
point in \(\Pi\).
\end{lem}
\begin{proof}
We can view the curves \(\tilde\gamma_+\) and \(\tilde\gamma_-\) as the graphs
of negative continuous function at the interval \((a,b)\). These functions
vanish at the point \(b\) and \(a\) respectively, hence they have at least one
intersection point \(\lambda_0\). Suppose, that there is another intersection
point \(\lambda_1\). We can assume that there are no other intersection points
at \(\tilde\gamma_+\) and \(\tilde\gamma_-\) between \(\lambda_0\) and
\(\lambda_1\). The parts of the curves \(\tilde\gamma_+\) and
\(\tilde\gamma_-\) between \(\lambda_0\) and \(\lambda_1\) form a closed
Jordan curve \(\tilde\gamma\), and the interior of \(\tilde\gamma\) is a
simply connected domain. Consider the function
\begin{align*}
	Q_0(\lambda)&=\int\limits_{\xi_0}^{\xi_{\lambda}}
	\sqrt{i(q(x)-\lambda)}\,dx,&\xi_0&=\xi_{\lambda_0},
\end{align*}
which is analytic in \(\Pi\). It follows from the definition of
\(\tilde\gamma_{\pm}\) that the harmonic function \(\Re Q_0(\lambda)\) vanishes
at both curves \(\tilde\gamma_+\) and \(\tilde\gamma_-\). In particular, \(\Re
Q_0(\lambda)=0\) for \(\lambda\in\tilde\gamma\). By virtue of the maximum
principle \(\Re Q_0(\lambda)\equiv 0\) in the interior of \(\tilde\gamma\).
This implies \(Q_0(\lambda)=\mathrm{const}\)
in \(\Pi\), although
\(Q(\lambda)\neq\mathrm{const}\). This contradiction ends the proof.
\end{proof}

Let \(\lambda_0\) be the intersection of \(\tilde\gamma_+\) and
\(\tilde\gamma_-\). It follows from~\eqref{eq:Qpm} that the curve
\(\tilde\gamma_{\infty}\) passes through the point \(\lambda_0\). By
Lemma~\ref{lem2:6} \(\tilde\gamma_{\infty}\) has no other intersection points
with \(\tilde\gamma_+\) and \(\tilde\gamma_-\). Denote by \(\gamma_+\),
\(\gamma_-\) and \(\gamma_{\infty}\) the parts of \(\tilde\gamma_+\),
\(\tilde\gamma_-\) and \(\tilde\gamma_{\infty}\) between the knot-point
\(\lambda_0\) and the points \(b\), \(a\) and \(-i\infty\), respectively. Set
\begin{equation}\label{eq:Gaa}
	\Gamma=\gamma_+\cup\gamma_-\cup\gamma_{\infty}.
\end{equation}
We shall show that \(\Gamma\) is the limit spectral graph of
problem~\eqref{eq2:1},~\eqref{eq2:2}. This implies the following: any point
\(\lambda\in\Gamma\) is an accumulation point for the eigenvalues, while the
complementary points \(\lambda\in\mathbb C\setminus\Gamma\) do not possess
this property. Moreover, in the end of this section we shall find explicit
formulae for the eigenvalue distribution along the curves \(\gamma_{\pm}\) and
\(\gamma_{\infty}\).
\begin{tm}\label{tm2:5}
Given \(\tau>0\) there exists an \(\varepsilon_0>0\) such that all the
eigenvalues of problem~\eqref{eq2:1},~\eqref{eq2:2} lie in the
\mbox{\(\tau\)-neigh}\-bour\-hood of \(\Gamma\), provided that
\(\varepsilon<\varepsilon_0\).
\end{tm}
\begin{proof}
We shall use in the sequel the results of asymptotic theory for ordinary
differential equation. A comprehensive exposition of this theory can be found
in books~\cite{O}, \cite{F}.

The function
\[
	S(z,\lambda)=\int\limits_{\xi_{\lambda}}^z\sqrt{i(q(\xi)-
	\lambda)}\,d\xi=0
\]
plays an important role in this theory. As before, here \(\xi_{\lambda}\) is
the root of the equation \(q(\xi)-\lambda=0\) from the domain \(G\). By
assumption the function \(q(z)\) is analytic in a neighbourhood of the segment
\([-1,1]\) and admits an analytic continuation in \(G\). Hence, there is a
number \(\delta>0\) such that \(q(z)\) is analytic in the domain
\(\Omega=G\cup U_{\delta}[-1,1]\), where \(U_{\delta}[-1,1]\) is the
\mbox{\(\delta\)-neigh}\-bour\-hood of \([-1,1]\) (see~Figure~\ref{fig4}).
Obviously, the function \(S(z,\lambda)\) is analytic with respect to the
variable \(\lambda\in\Pi\) and continuous for \(\lambda\in\overline{\Pi}\), and
locally analytic with respect to the variable \(z\in G\) with the branching
point \(\xi_{\lambda}\).

Fix a number \(\lambda=c\in(-1,1)\) and chose the branch of this function such
that \(\Re S(1,c)>0\). For other values \(\lambda\in\overline{\Pi}\) the
branch of \(S(z,\lambda)\) is chosen by the following condition: the values
\(S(1,c)\) and \(S(1,\lambda)\) are analytically
connected.

\begin{figure}[t]
\setlength{\unitlength}{0.0005489in}
\begingroup\makeatletter\ifx\SetFigFont\undefined%
\gdef\SetFigFont#1#2#3#4#5{%
  \reset@font\fontsize{#1}{#2pt}%
  \fontfamily{#3}\fontseries{#4}\fontshape{#5}%
  \selectfont}%
\fi\endgroup%
{\renewcommand{\dashlinestretch}{30}
\begin{picture}(6185,4414)(0,-10)
\drawline(1092.000,3982.000)(1018.861,3977.740)(946.712,3965.018)
	(876.527,3944.006)(809.257,3914.989)(745.809,3878.357)
	(687.044,3834.608)(633.755,3784.332)(586.662,3728.210)
	(546.404,3667.000)(513.524,3601.530)(488.467,3532.686)
	(471.571,3461.398)(463.066,3388.631)(463.066,3315.369)
	(471.571,3242.602)(488.467,3171.314)(513.524,3102.470)
	(546.404,3037.000)(586.662,2975.790)(633.755,2919.668)
	(687.044,2869.392)(745.809,2825.643)(809.257,2789.011)
	(876.527,2759.994)(946.712,2738.982)(1018.861,2726.260)
	(1092.000,2722.000)
\drawline(5592.000,2722.000)(5664.066,2731.540)(5734.558,2749.295)
	(5802.544,2775.031)(5867.123,2808.407)(5927.441,2848.981)
	(5982.698,2896.215)(6032.163,2949.485)(6075.180,3008.085)
	(6111.181,3071.239)(6139.689,3138.110)(6160.325,3207.813)
	(6172.817,3279.426)(6177.000,3352.000)(6172.817,3424.574)
	(6160.325,3496.187)(6139.689,3565.890)(6111.181,3632.761)
	(6075.180,3695.915)(6032.163,3754.515)(5982.698,3807.785)
	(5927.441,3855.019)(5867.123,3895.593)(5802.544,3928.969)
	(5734.558,3954.705)(5664.066,3972.460)(5592.000,3982.000)
\put(3207,2812){\circle*{90}}
\put(2577,3352){\circle*{90}}
\put(4062,3352){\circle*{90}}
\put(1092,3352){\circle{90}}
\put(5592,3352){\circle{90}}
\drawline(1092,3352)(5592,3352)
\drawline(1092,3982)(5592,3982)
\drawline(1092,2722)(1091,2720)(1090,2716)
	(1088,2708)(1085,2696)(1080,2679)
	(1074,2656)(1066,2628)(1056,2594)
	(1045,2555)(1033,2512)(1019,2464)
	(1005,2413)(990,2360)(974,2305)
	(957,2249)(941,2193)(925,2138)
	(909,2083)(893,2030)(877,1979)
	(862,1929)(847,1881)(832,1836)
	(818,1792)(804,1750)(791,1709)
	(777,1670)(764,1633)(751,1597)
	(738,1561)(725,1527)(712,1493)
	(699,1460)(686,1427)(672,1394)
	(658,1362)(644,1329)(629,1297)
	(614,1264)(599,1230)(583,1196)
	(566,1161)(548,1126)(530,1089)
	(510,1051)(489,1011)(468,970)
	(445,927)(421,883)(396,837)
	(370,789)(343,740)(315,690)
	(287,640)(258,588)(229,537)
	(201,487)(173,438)(147,392)
	(122,348)(99,309)(79,273)
	(61,242)(46,217)(35,196)
	(26,180)(19,169)(15,162)
	(13,159)(12,157)
\drawline(5592,2722)(5592,2720)(5590,2716)
	(5588,2708)(5585,2696)(5581,2678)
	(5575,2655)(5567,2625)(5558,2590)
	(5548,2549)(5536,2503)(5523,2452)
	(5508,2397)(5493,2339)(5478,2279)
	(5462,2218)(5445,2155)(5429,2093)
	(5413,2031)(5397,1971)(5381,1912)
	(5366,1855)(5351,1800)(5337,1747)
	(5323,1697)(5309,1648)(5297,1602)
	(5284,1558)(5272,1515)(5260,1475)
	(5249,1436)(5237,1398)(5226,1362)
	(5215,1326)(5204,1292)(5194,1258)
	(5183,1225)(5172,1192)(5160,1156)
	(5148,1120)(5135,1084)(5122,1049)
	(5109,1013)(5096,977)(5082,940)
	(5068,902)(5053,864)(5037,824)
	(5020,783)(5003,741)(4985,697)
	(4966,652)(4946,605)(4926,557)
	(4905,508)(4884,459)(4862,409)
	(4841,359)(4820,311)(4799,264)
	(4780,220)(4762,180)(4746,143)
	(4732,112)(4720,85)(4710,63)
	(4703,47)(4698,35)(4695,28)
	(4693,24)(4692,22)
\thicklines
\drawline(5232,3982)(5230,3981)(5226,3979)
	(5219,3975)(5207,3970)(5190,3962)
	(5168,3951)(5141,3938)(5109,3922)
	(5073,3904)(5032,3884)(4988,3863)
	(4942,3840)(4894,3816)(4846,3792)
	(4797,3768)(4748,3744)(4701,3720)
	(4655,3697)(4610,3675)(4567,3653)
	(4526,3632)(4487,3612)(4450,3593)
	(4414,3575)(4380,3557)(4347,3539)
	(4316,3523)(4286,3506)(4256,3490)
	(4228,3474)(4200,3458)(4172,3443)
	(4145,3427)(4116,3410)(4087,3393)
	(4058,3376)(4029,3358)(3999,3340)
	(3969,3322)(3939,3303)(3907,3283)
	(3874,3261)(3840,3239)(3805,3216)
	(3768,3192)(3729,3167)(3689,3140)
	(3648,3112)(3606,3084)(3563,3055)
	(3520,3026)(3477,2997)(3435,2968)
	(3395,2941)(3357,2915)(3323,2892)
	(3293,2872)(3268,2854)(3247,2840)
	(3231,2829)(3220,2821)(3213,2816)
	(3209,2813)(3207,2812)
\drawline(3207,2812)(3207,2810)(3206,2806)
	(3204,2799)(3201,2787)(3196,2770)
	(3191,2748)(3184,2720)(3175,2687)
	(3166,2649)(3155,2606)(3143,2560)
	(3130,2510)(3116,2457)(3102,2403)
	(3087,2348)(3073,2293)(3058,2238)
	(3044,2184)(3029,2131)(3015,2080)
	(3001,2031)(2988,1983)(2974,1937)
	(2962,1892)(2949,1850)(2936,1809)
	(2924,1769)(2912,1730)(2900,1692)
	(2887,1656)(2875,1619)(2863,1583)
	(2850,1548)(2838,1512)(2825,1477)
	(2812,1443)(2799,1409)(2786,1375)
	(2772,1340)(2758,1305)(2743,1269)
	(2728,1232)(2712,1193)(2695,1154)
	(2677,1113)(2658,1070)(2639,1026)
	(2618,980)(2597,932)(2574,882)
	(2550,830)(2526,776)(2500,721)
	(2474,664)(2447,606)(2420,548)
	(2393,490)(2366,433)(2340,377)
	(2315,323)(2291,272)(2268,225)
	(2248,182)(2230,144)(2215,111)
	(2202,84)(2191,63)(2184,46)
	(2178,35)(2175,28)(2173,24)(2172,22)
\thinlines
\dashline{150.000}(5232,3982)
	(5608,4170)(5917,4324)(6042,4387)
\dashline{150.000}(507,4072)
	(754,3675)(803,3271)
	(653,2862)(507,2632)
\thicklines
\drawline(3207,2812)(3205,2814)(3201,2819)
	(3193,2828)(3181,2842)(3166,2859)
	(3149,2879)(3129,2901)(3109,2924)
	(3088,2947)(3068,2969)(3049,2990)
	(3031,3009)(3014,3027)(2999,3044)
	(2984,3059)(2970,3073)(2956,3087)
	(2943,3100)(2930,3112)(2916,3124)
	(2902,3137)(2888,3149)(2873,3161)
	(2858,3174)(2842,3187)(2825,3200)
	(2808,3213)(2790,3226)(2771,3240)
	(2752,3253)(2733,3266)(2713,3279)
	(2693,3292)(2673,3305)(2653,3318)
	(2632,3330)(2612,3342)(2591,3355)
	(2570,3367)(2551,3377)(2532,3388)
	(2513,3399)(2492,3410)(2471,3422)
	(2449,3434)(2425,3446)(2401,3459)
	(2377,3471)(2351,3485)(2325,3498)
	(2299,3512)(2272,3525)(2245,3539)
	(2218,3552)(2191,3566)(2164,3579)
	(2137,3592)(2111,3605)(2085,3618)
	(2060,3631)(2034,3643)(2009,3655)
	(1985,3667)(1961,3678)(1938,3689)
	(1915,3701)(1890,3712)(1865,3724)
	(1839,3737)(1811,3750)(1782,3764)
	(1751,3778)(1718,3794)(1684,3810)
	(1648,3827)(1610,3845)(1571,3863)
	(1533,3881)(1494,3899)(1458,3916)
	(1424,3932)(1394,3946)(1369,3958)
	(1349,3967)(1334,3974)(1325,3978)
	(1319,3981)(1317,3982)
\thinlines
\dashline{150.000}(1317,3982)(687,4252)
\put(1137,2992){\makebox(0,0)[lb]{\(-1\)}}
\put(5367,2992){\makebox(0,0)[lb]{\(+1\)}}
\put(3342,2452){\makebox(0,0)[lb]{\(\xi_{\lambda}\)}}
\put(4152,3037){\makebox(0,0)[lb]{\(c_+\)}}
\put(2352,3082){\makebox(0,0)[lb]{\(c_-\)}}
\put(2257,3677){\makebox(0,0)[lb]{\(l_2\)}}
\put(4257,3677){\makebox(0,0)[lb]{\(l_1\)}}
\put(2387,1282){\makebox(0,0)[lb]{\(l_3\)}}
\put(3522,1597){\makebox(0,0)[lb]{\(G\)}}
\end{picture}
}
\caption{}\label{fig4}
\end{figure}

For a fixed \(\lambda\in\Pi\) the set
\[
	\{z\;\vline\;\Re S(z,\lambda)=0\},
\]
determines the lines in the \mbox{\(z\)-pla}\-ne which are called the Stokes
lines. The point \(\xi_{\lambda}\) belongs to this set and three lines come
out from this point. Conventionally, we will use the terms \textit{the right,
the left and the lower Stokes lines}. Although these terms are not defined
rigorously, the identification will be clear from the context. To clarify the
situation, we remark that for a fixed \(\lambda\in\Pi\) the domain \(G\) may
contain some other Stokes lines, besides the Stokes complex with the
knot-point \(\xi_{\lambda}\) (such a line is depicted near the point \(-1\) in
Figure~\ref{fig4}). In this connection the following assertion plays an
important role.
\begin{lem}\label{lem2:5}
Let \(\lambda\in\Pi\setminus\left\{\gamma_+\cup\gamma_-\right\}\), where
\(\gamma_{\pm}\) are defined in~\eqref{eq:Gaa}. Consider all the cases
(see~Figure~\ref{fig3}):
\begin{enumerate}
\item The point \(\lambda\) lies above both curves \(\tilde\gamma_+\) è
\(\tilde\gamma_-\);
\item \(\lambda\) lies under the curve \(\tilde\gamma_-\), but above the curve
\(\tilde\gamma_+\);
\item \(\lambda\in\tilde\gamma_+\setminus\gamma_+\), i.~e. \(\lambda\) belongs
to \(\tilde\gamma_+\), but lie under the curve \(\tilde\gamma_-\);
\item \(\lambda\) lies under the curve \(\tilde\gamma_+\), but above
\(\tilde\gamma_-\);
\item \(\lambda\in\tilde\gamma_-\setminus\gamma_-\), i.~e. \(\lambda\) belongs
to \(\tilde\gamma_-\) and lies under \(\tilde\gamma_+\);
\item \(\lambda\) lies under both curves
\(\tilde\gamma_+\) and
\(\tilde\gamma_-\).
\end{enumerate}
Then, in the first case the left and the right Stokes lines intersect the
interval \((-1,1)\) in points \(c^-=c^-(\lambda)\) and \(c^+=c^+(\lambda)\),
and no other Stokes line intersect the segment \([-1,1]\). In the second case
only the right Stokes line intersect the interval \((-1,1)\) in a point
\(c^+=c^+(\lambda)\) and no other Stokes lines intersect the segment
\([-1,1]\). In the third case the right Stokes line intersect the point \(1\)
and no other Stokes lines intersect \([-1,1]\). The fourth and the fifth cases
are similar to the second and the third ones, only the roles of the points
\(-1\) and \(1\) are changed. Finally, in the sixth case no Stokes lines
intersect \([-1,1]\).
\end{lem}
\begin{proof}
The branch of the function \(S(z,\lambda)\) is chosen by the condition \(\Re
S(1,c)>0\) for some (and hence for all) \(c\in(-1,1)\). Viewing \(S(z,c)\) as
an analytic function on \(z\in G\cup[-1,c)\cup(c,1]\), we find that \(\Re
S(z,c)>0\) for \(c<z\leqslant 1\) and \(\Re S(z,c)<0\) for \(-1\leqslant
z<c\). It is proved in Lemma~\ref{lem2:1} that the function \(\Re
S(1,c-it)=\Re Q^+(c-it)\) decreases monotonously as \(t\) runs from \(0\) to
\(+\infty\) and vanishes in the only point. The Stokes lines continuously
depend on \(\lambda\), hence, for small values \(t\in[0,t_0]\) these lines
intersect the interval \((-1,1)\) in points \(c^-(t)\), \(c^+(t)\) which are
close to the point \(c\) (see~Figure~\ref{fig4}). For all \(z\in(c^+,1]\) we
have
\[
	\Re S(z,c-it)=\Re\left(\int\limits_{\xi_{\lambda}}^{c^+}+
	\int\limits_{c^+}^z\sqrt{i(q(x)-\lambda)}\,dx\right)=
	\Re\int\limits_{c^+}^z\sqrt{i(q(x)-\lambda)}\,dx>0
\]
since the point \(c^+\) lies on the Stokes line and the values of
\(q(x)-\lambda\) belong to the first quadrant of the complex plane for
\(\lambda=c-it\) and \(x\geqslant c\). This inequality shows that no Stokes
lines intersect the set \((c^+,1]\). Similarly, no Stokes lines intersect the
set \([-1,c^-)\), and the interval \((c^-,c^+)\).

Further, the function
\[
	\Re Q^+(c-it)=\Re S(1,c-it)=\Re\int\limits_{c^+}^1
	\sqrt{i(q(x)-\lambda)}\,dx
\]
decreases monotonously as \(t\) grows from \(0\). Therefore, the point \(c^+=
c^+(t)\) moves to the right and reaches the point \(1\) as \(\lambda=c-it\)
comes to the curve \(\tilde\gamma_+\). Similarly, the point \(c^-(t)\) moves
to the left and reaches to the point \(-1\) as \(\lambda=c-it\) comes the
curve \(\tilde\gamma_-\). This analysis makes obvious all other assertions of
Lemma~\ref{lem2:5}.
\end{proof}

Now, let us recall an important concept of \textit{canonical domain} for
equation~\eqref{eq2:1}. A domain \(\Omega_{\lambda}\) in the
\mbox{\(z\)-pla}\-ne is called canonical if the function \(S(z,\lambda)\)
is univalent in \(\Omega_{\lambda}\) (we do not define here maximal canonical
domains). It follows easily from the definition that any domain not containing
the points of the Stokes graph is canonical. Moreover, \(\Re S(z,\lambda)\)
preserves the sign for \(z\) belonging to such a domain. This fact implies
that any domain containing only one Stokes line remains to be canonical.
\begin{lem}\label{lem2:6+}
Given \(\lambda\in\Pi\setminus(\gamma_+\cup\gamma_-)\) there exist a path
connecting the points \(\pm 1\) and a canonical domain \(\Omega_{\lambda}\)
which entirely contains this path.
\end{lem}
\begin{proof}
Consider, for instance, the case when the point \(\lambda\) lies above both
curves \(\tilde\gamma_+\) è \(\tilde\gamma_-\). Recall the notation
\(\Omega=G\cup U_{\delta}[-1,1]\) and consider the domain
\(\Omega\setminus\Omega_{\lambda}^+\), where \(\Omega_{\lambda}^+\) is the
domain in \(\Omega\) bounded by the left and the right Stokes lines and
containing the interval \((c^-,c^+)\subset(-1,1)\) (see~Figure~\ref{fig4}).
The domain \(\Omega\setminus\Omega_{\lambda}^+\) contains the lower Stokes line
outgoing from \(\xi_{\lambda}\) and, probably, some other Stokes lines.
However, these other lines do not intersect the lines outgoing from the point
\(\xi_{\lambda}\) (this is general property), and do not intersect the sets
\([-1,c^-)\) and \((c^+,1]\) by virtue of Lemma~\ref{lem2:5}. Hence, there is
a path in \(\Omega\setminus\Omega_{\lambda}^+\) connecting the points \(\pm
1\) which intersects only the lower Stokes line, and there exists a
neighbourhood of this path containing no Stokes lines but a part of the lower
one. Such a domain is canonical (as we have noted before). All the other cases
of disposition of a point \(\lambda\) (see~Lemma~\ref{lem2:5}) can be treated
analogously. Lemma is proved.
\end{proof}

The following well-known fact will be essentially used in the sequel.
\begin{lem}\label{lem2:77}
Given fixed \(\lambda\in\Pi\) equation~\eqref{eq2:1} possesses two linear
independent solution of the form
\begin{equation}\label{eq2:5}
	v_{\pm}(z,\lambda)=\dfrac{1}{[i(q(z)-\lambda)]^{1/4}}
	e^{\pm\varepsilon^{-1}S(z,\lambda)}\left(1+O_{\pm}(\varepsilon)
	\right),
\end{equation}
where the function \(O_{\pm}\) are subject the estimate
\[
	\left|O_{\pm}(\varepsilon)\right|\leqslant C\varepsilon
\]
with a constant \(C\) not depending on \(z\), as \(z\) varies on a compact
\(K\) belonging to a canonical domain \(\Omega_{\lambda}\). Moreover, if
\(K\subset\Omega_{\lambda}\) as \(\lambda\) varies on a compact \(K'\) in the
\mbox{\(\lambda\)-pla}\-ne, then the estimate holds with a constant \(C\) not
depending on \(z\in K\) and \(\lambda\in K'\).
\end{lem}
\begin{proof}
See~\cite{O,F}.
\end{proof}

Now, we start a direct proof of Theorem~\ref{tm2:5}. Fix an arbitrary number
\(\tau>0\) and denote by \(\Gamma_{\tau}\) the
\mbox{\(\tau\)-neigh}\-bour\-hood of the limit spectral graph \(\Gamma\). We
have to consider six different cases mentioned in Lemma~\ref{lem2:5} for the
disposition of the point \(\lambda\) with respect to the curves
\(\tilde\gamma_+\) and \(\tilde\gamma_-\). Consider, for instance the first
case, when \(\lambda\) lies under both curves \(\tilde\gamma_+\) and
\(\tilde\gamma_-\). The characteristic determinant of
problem~\eqref{eq2:1},~\eqref{eq2:2} compiled from fundamental
solutions~\eqref{eq2:5} has the representation
\begin{equation}\label{eq2:6}
	\Delta(\lambda)=\begin{vmatrix}v^+(1,\lambda)&v^+(-1,\lambda)\\
	v^-(1,\lambda)&v^-(-1,\lambda)\end{vmatrix}=T(\lambda)
	\left(e^{\varepsilon^{-1}(S(1,\lambda)-S(-1,\lambda))}[1]-
	e^{-\varepsilon^{-1}(S(1,\lambda)-S(-1,\lambda))}[1]\right),
\end{equation}
where the function
\[
	T(\lambda)=(i(q(1)-\lambda)^{-1/4}(i(q(-1)-\lambda))^{-1/4}
\]
does not vanish in \(\Pi\). For abriviation we use the Birkhoff notation
\([1]=1+O(\varepsilon)\). It follows from representation~\eqref{eq2:6} that
\(\Delta(\lambda)\neq 0\) if
\begin{equation}\label{eq2:7}
	\Re(S(1,\lambda)-S(-1,\lambda))>0
\end{equation}
and \(\varepsilon<\varepsilon_0=\varepsilon_0(\lambda)\). 

Lemma~\ref{lem2:77} asserts that the obtained asymptotic representation for
\(\Delta(\lambda)\) is valid, if the points \(\pm 1\) can be connected by a
path lying in a canonical domain of equation~\eqref{eq2:1}.
Lemma~\ref{lem2:6+} guarantees the existence of such a path
\(\gamma_{\lambda}\) and a canonical domain \(\Omega_{\lambda}\). Moreover,
the points \(+1\) and \(-1\) is this canonical domain are separated by a lower
Stokes line. Recall that the function \(\Re S(z,\lambda)\) changes the sign
passing through the Stokes line. We fixed the branch of the function
\(S(z,\lambda)\) by the condition \(\Re S(1,c)>0\) for \(c\in(-1,1)\), and
this implies \(\Re S(1,\lambda)>0\) if \(\lambda\) lies under the curves
\(\tilde\gamma_+\) and \(\tilde\gamma_-\). Therefore, \(\Re S(-1,\lambda)<0\)
and inequality~\eqref{eq2:7} holds. This inequality remains valid in a
neighbourhood \(U_{\lambda}\) of the point \(\lambda\), by continuity. The
neighbourhood \(U_{\lambda}\) can be chosen such small that a path \(\gamma\)
connected the points \(1\) and \(-1\) does not intersect the Stokes lines of
the Stokes graph \(\Gamma_{\mu}\) for all \(\mu\in U_{\lambda}\) with
exception of the lower lines. Therefore, the canonical domain
\(\Omega_{\lambda}\) containing \(\gamma\) can be chosen such small that it
remains to be canonical for all \(\mu\in U_{\lambda}\), i.~e. we can set
\(\Omega_{\mu}=\Omega_{\lambda}\). By virtue of Lemma~\ref{lem2:77} asymptotic
representation~\eqref{eq2:6} for \(\Delta(\lambda)\) remains to be valid in
the whole neighbourhood \(U_{\lambda}\) of the point \(\lambda\), and the
remainders in the representations \([1]=1+O(\varepsilon)\) in~\eqref{eq2:6}
can be estimated by \(C\varepsilon\) with a constant \(C\) not depending on
\(\mu\in U_{\lambda}\).

Fix a number \(R\gg 1\) and denote by \(\overline{\Pi}_R\) the intersection of
the semistrip \(\overline{\Pi}\) with the closed circle of radius \(R\) with
center at the origin. For sufficiently small \(\tau\) the set
\(\overline{\Pi}_R\setminus\Gamma_{\tau}\) consists of three compacts, say, the
upper, the left and the right ones. With each point \(\lambda\) belonging to
the upper compact \(K^+\) we associate a neighbourhood \(U_{\lambda}\), that
was constructed above, and from the cover of \(K^+\) by these neighbourhoods
take a finite subcover. This procedure allows to obtain
representation~\eqref{eq2:6} for all \(\lambda\in K^+\), moreover,
\begin{align*}
	\Re(S(1,\lambda)-S(-1,\lambda))&>c>0,&\lambda&\in K^+,
\end{align*}
with a constant \(c\) dependent only on \(\tau\),
and the estimates
\(|O(\varepsilon)|<C\varepsilon\) hold for the remainders in the
representations \([1]=1+O(\varepsilon)\) with a constant \(C\)
dependent also
only on \(\lambda\). Then, we obtain \(\Delta(\lambda)\neq 0\) for all
\(\lambda\in K^+\) if \(\varepsilon<\varepsilon_0\) and
\(\varepsilon_0=\varepsilon_0(\tau)\) is sufficiently small.

Analogously, we can prove the absence of the zeros of \(\Delta(\lambda)\) in
the left and the right compacts. It is left to show that the choice of a
\mbox{\(\tau\)-neigh}\-bour\-hood of the curve \(\gamma_{\infty}\) can be
realized independently on \(R\), i.~e. the eigenvalues with large moduli do
not leave the \mbox{\(\tau\)-neigh}\-bour\-hood of the curve
\(\gamma_{\infty}\), vice versa, asymptotically they lie more close to this
curve. This fact can be proved in the same way as in Lemma~4.4 of
paper~\cite{ShT} (see Remark~4.1 in ~\cite{ShT}). Theorem is proved.
\end{proof}

Next, we shall obtain an additional information on the eigenvalue behaviour
near the curves of the limit spectral graph. For this purposes we need the
concept of counting function of the zeros of holomorphic functions along the
curves. 

Let \(\gamma=\gamma(t)\), \(t\in[0,1]\), be an oriented smooth curve in the
complex plane \(\mathbb C\) with end points \(z_0\) and \(z_1\) (the value
\(\infty\) for these points is admitted). One can order points on the curve as
follows: \(\lambda_1\prec\lambda_2\) if
\(\lambda_j=\gamma(t_j)\) and \(t_1<t_2\). Denote by
\(\gamma_{\tau}(\lambda_1,\lambda_2)\) a curvelinear strip of width \(2\tau\)
containing \(\gamma\) as the middle line and having, as the lateral sides, the
segments perpendicular to \(\gamma\) at the points \(\lambda_1\) and
\(\lambda_2\). Let a function \(F(z)\) be holomorphic in
\(\gamma_{\tau}(z_0,z_1)\). Fix a point \(\lambda_1\in\gamma\) and denote by
\(n(\lambda_1,\lambda)\) the number of zeros of the function \(F(z)\) in
\(\gamma_{\tau}(\lambda_1,\lambda)\), if \(\lambda_1\prec\lambda\). For
\(\lambda\prec\lambda_1\) define
\(n(\lambda_1,\lambda)=-n(\lambda,\lambda_1)\). Now, define
\(N(\lambda)=n(\lambda_1,\lambda)+C\), where \(C\) is an arbitrary constant, as
the \textit{zero counting function of} \(F(z)\) in a
\mbox{\(\tau\)-neigh}\-bour\-hood of the curve \(\gamma\) (or along the curve
\(\gamma\)). Since the eigenvalues of the problem in question are the zeros of
the entire function \(\Delta(\lambda)\), we nay speak in the same context on
the eigenvalue counting functions along the curves.
\begin{tm}\label{tm2:2}
Fix a small number \(\delta>0\) and denote by \(\mu_k^+\), \(\mu_k^-\) and
\(\mu_k\) the solutions of the equations
\begin{align*}
	i\int\limits_{\xi_{\lambda}}^1\sqrt{i(q(\xi)-\lambda)}\,d\xi&=
	\varepsilon
	\pi(k-1/4),&k&\in\mathbb Z,\\
	i\int\limits_{\xi_{\lambda}}^{-1}\sqrt{i(q(\xi)-\lambda)}\,d\xi&=
	\varepsilon
	\pi(k-1/4),&k&\in\mathbb Z,\\
	i\int\limits_{-1}^1\sqrt{i(q(\xi)-\lambda)}\,d\xi&=
	\varepsilon\pi k,&k&\in\mathbb Z,
\end{align*}
lying on the curves \(\gamma_+\), \(\gamma_-\) and \(\gamma_{\infty}\) (note,
that left hand-sides take real values along the corresponding curves). Chose
the indices \(p_{\pm}\), \(m_{\pm}\) and \(s_0\) such that
\(\{\mu_k^+\}_{p_+}^{m_+}\),
\(\{\mu_k^-\}_{p_-}^{m_-}\) and
\(\{\mu_k\}_{s_0}^{\infty}\) are all solutions
of these equations on the curves \(\gamma_+\), \(\gamma_-\) and
\(\gamma_{\infty}\), respectively, lying outside the
\mbox{\(\delta\)-neigh}\-bour\-hoods of the points \(a\), \(b\) and the
knot-point \(\lambda_0\). Then, there exists a number \(C\) depending only on
\(\delta\), such that all the eigenvalues lie in the
\mbox{\(\delta\)-neigh}\-bour\-hoods \(U_{\delta}(a)\), \(U_{\delta}(b)\) and
\(U_{\delta}(\lambda_0)\), and in the circles of the radius \(C\varepsilon^2\)
centered at the points \(\{\mu_k^+\}_{p_+-1}^{m_++1}\),
\(\{\mu_k^-\}_{p_--1}^{m_-+1}\) and \(\{\mu_k\}_{s_0-1}^{\infty}\). All these
circles contain only one
eigenvalue. The eigenvalue counting functions along the curves \(\gamma_+\),
\(\gamma_-\) and \(\gamma_{\infty}\) have representations
\begin{align*}
	N(\lambda)&=\dfrac{1}{i\pi\varepsilon}Q^+(\lambda)+O(1),&\text{if }
	\lambda\in\gamma_+,\\
	N(\lambda)&=\dfrac{1}{i\pi\varepsilon}Q^-(\lambda)+O(1),&\text{if }
	\lambda\in\gamma_-,\\
	N(\lambda)&=\dfrac{1}{i\pi\varepsilon}Q(\lambda)+O(1),&\text{if }
	\lambda\in\gamma_{\infty}.
\end{align*}
The remainders in these formulae are estimated by a constant not
dependent on
\(\lambda\) and \(\varepsilon\), if \(\lambda\) lies outside the
neighbourhoods \(U_{\delta}(a)\), \(U_{\delta}(b)\) and
\(U_{\delta}(\lambda_0)\).
\end{tm}
\begin{proof}
Consider, for example, the curve \(\gamma_+\). We shall make use of the
transmission formulae for asymptotic representations of solutions in
neighbouring canonical domains. Below we formulate the corresponding result in
a convenient form, viewing in mind our concrete
problem.

\begin{figure}[t]
\setlength{\unitlength}{0.00067489in}
\begingroup\makeatletter\ifx\SetFigFont\undefined%
\gdef\SetFigFont#1#2#3#4#5{%
  \reset@font\fontsize{#1}{#2pt}%
  \fontfamily{#3}\fontseries{#4}\fontshape{#5}%
  \selectfont}%
\fi\endgroup%
{\renewcommand{\dashlinestretch}{30}
\begin{picture}(4534,2444)(0,-10)
\put(462,1957){\blacken\ellipse{90}{90}}
\put(462,1957){\ellipse{90}{90}}
\put(4062,1957){\blacken\ellipse{90}{90}}
\put(4062,1957){\ellipse{90}{90}}
\put(2712,1192){\blacken\ellipse{90}{90}}
\put(2712,1192){\ellipse{90}{90}}
\path(462,1957)(4062,1957)
\path(462,2407)(4062,2407)
\path(462,2407)(4062,2407)
\path(462,2407)(461,2407)(456,2406)
	(445,2403)(428,2398)(408,2393)
	(386,2387)(366,2382)(348,2377)
	(333,2373)(319,2369)(308,2365)
	(297,2362)(285,2358)(273,2353)
	(262,2349)(250,2344)(239,2338)
	(228,2333)(218,2327)(209,2321)
	(200,2316)(192,2309)(184,2303)
	(175,2296)(166,2288)(157,2280)
	(148,2271)(139,2262)(131,2253)
	(123,2244)(116,2235)(110,2227)
	(103,2219)(98,2210)(92,2201)
	(86,2191)(81,2180)(75,2169)
	(70,2157)(66,2146)(61,2134)
	(57,2122)(54,2111)(50,2100)
	(47,2087)(43,2074)(40,2060)
	(37,2045)(34,2030)(31,2014)
	(29,1999)(28,1985)(27,1971)
	(27,1957)(27,1943)(28,1929)
	(29,1915)(31,1900)(34,1884)
	(37,1869)(40,1854)(43,1840)
	(47,1827)(50,1814)(54,1803)
	(57,1792)(61,1780)(66,1768)
	(70,1757)(75,1745)(81,1734)
	(86,1723)(92,1713)(98,1704)
	(103,1695)(110,1687)(116,1679)
	(123,1670)(131,1661)(139,1652)
	(148,1643)(157,1634)(166,1626)
	(175,1618)(184,1611)(192,1604)
	(200,1598)(209,1593)(218,1587)
	(228,1581)(239,1576)(250,1570)
	(262,1565)(273,1561)(285,1556)
	(297,1552)(308,1549)(319,1545)
	(333,1541)(348,1537)(366,1532)
	(386,1527)(408,1521)(428,1516)
	(445,1511)(456,1508)(461,1507)(462,1507)
\path(4062,2407)(4063,2407)(4068,2406)
	(4079,2403)(4096,2398)(4116,2393)
	(4138,2387)(4158,2382)(4176,2377)
	(4191,2373)(4205,2369)(4216,2365)
	(4227,2362)(4239,2358)(4251,2353)
	(4262,2349)(4274,2344)(4285,2338)
	(4296,2333)(4306,2327)(4315,2321)
	(4324,2316)(4332,2309)(4340,2303)
	(4349,2296)(4358,2288)(4367,2280)
	(4376,2271)(4385,2262)(4393,2253)
	(4401,2244)(4408,2235)(4415,2227)
	(4421,2219)(4426,2210)(4432,2201)
	(4438,2191)(4443,2180)(4449,2169)
	(4454,2157)(4458,2146)(4463,2134)
	(4467,2122)(4470,2111)(4474,2100)
	(4477,2087)(4481,2074)(4484,2060)
	(4487,2045)(4490,2030)(4493,2014)
	(4495,1999)(4496,1985)(4497,1971)
	(4497,1957)(4497,1943)(4496,1929)
	(4495,1915)(4493,1900)(4490,1884)
	(4487,1869)(4484,1854)(4481,1840)
	(4477,1827)(4474,1814)(4470,1803)
	(4467,1792)(4463,1780)(4458,1768)
	(4454,1757)(4449,1745)(4443,1734)
	(4438,1723)(4432,1713)(4426,1704)
	(4421,1695)(4415,1687)(4408,1679)
	(4401,1670)(4393,1661)(4385,1652)
	(4376,1643)(4367,1634)(4358,1626)
	(4349,1618)(4340,1611)(4332,1604)
	(4324,1598)(4315,1593)(4306,1587)
	(4296,1581)(4285,1576)(4274,1570)
	(4262,1565)(4251,1561)(4239,1556)
	(4227,1552)(4216,1549)(4205,1545)
	(4191,1541)(4176,1537)(4158,1532)
	(4138,1527)(4116,1521)(4096,1516)
	(4079,1511)(4068,1508)(4063,1507)(4062,1507)
\path(462,1507)(462,1505)(461,1499)
	(460,1489)(458,1474)(456,1454)
	(453,1429)(450,1399)(446,1367)
	(442,1332)(437,1296)(432,1260)
	(427,1225)(422,1190)(416,1157)
	(411,1125)(405,1095)(400,1066)
	(394,1037)(387,1010)(380,982)
	(373,955)(365,928)(357,899)
	(350,877)(342,853)(334,829)
	(326,804)(317,778)(307,750)
	(296,722)(285,691)(272,659)
	(259,624)(245,588)(230,549)
	(213,509)(196,466)(178,422)
	(160,377)(141,331)(122,286)
	(104,242)(86,200)(70,161)
	(56,126)(43,96)(33,71)
	(25,52)(19,38)(15,29)
	(13,24)(12,22)
\path(4062,1507)(4062,1505)(4061,1499)
	(4060,1489)(4058,1474)(4056,1454)
	(4053,1429)(4050,1399)(4046,1367)
	(4042,1332)(4037,1296)(4032,1260)
	(4027,1225)(4022,1190)(4016,1157)
	(4011,1125)(4005,1095)(4000,1066)
	(3994,1037)(3987,1010)(3980,982)
	(3973,955)(3965,928)(3957,899)
	(3950,877)(3942,853)(3934,829)
	(3926,804)(3917,778)(3907,750)
	(3896,722)(3885,691)(3872,659)
	(3859,624)(3845,588)(3830,549)
	(3813,509)(3796,466)(3778,422)
	(3760,377)(3741,331)(3722,286)
	(3704,242)(3686,200)(3670,161)
	(3656,126)(3643,96)(3633,71)
	(3625,52)(3619,38)(3615,29)
	(3613,24)(3612,22)
\thicklines
\path(2712,1192)(2711,1189)(2709,1184)
	(2706,1174)(2701,1158)(2694,1137)
	(2686,1111)(2676,1079)(2664,1044)
	(2652,1006)(2639,966)(2626,926)
	(2613,886)(2600,847)(2587,810)
	(2575,775)(2564,741)(2553,710)
	(2543,681)(2533,654)(2524,628)
	(2515,604)(2506,581)(2497,559)
	(2488,538)(2480,517)(2469,493)
	(2459,469)(2448,445)(2436,421)
	(2424,397)(2412,372)(2398,345)
	(2383,317)(2368,288)(2351,258)
	(2333,226)(2315,194)(2297,161)
	(2280,130)(2263,102)(2249,77)
	(2237,56)(2228,41)(2222,30)
	(2219,25)(2217,22)
\path(2712,1192)(2710,1193)(2706,1196)
	(2698,1202)(2687,1210)(2671,1221)
	(2650,1236)(2627,1253)(2600,1273)
	(2571,1294)(2541,1317)(2510,1340)
	(2479,1363)(2449,1386)(2421,1409)
	(2394,1431)(2368,1453)(2344,1473)
	(2321,1494)(2299,1514)(2279,1533)
	(2259,1553)(2240,1573)(2222,1593)
	(2205,1613)(2187,1634)(2172,1653)
	(2157,1673)(2142,1693)(2126,1714)
	(2111,1736)(2095,1760)(2079,1785)
	(2062,1811)(2044,1840)(2025,1870)
	(2006,1902)(1986,1936)(1965,1972)
	(1943,2009)(1921,2049)(1898,2089)
	(1875,2130)(1852,2170)(1830,2210)
	(1809,2247)(1790,2282)(1773,2314)
	(1758,2341)(1746,2363)(1737,2380)
	(1730,2393)(1725,2401)(1723,2405)(1722,2407)
\path(2712,1192)(2714,1193)(2718,1194)
	(2727,1197)(2739,1201)(2757,1206)
	(2781,1213)(2809,1222)(2843,1233)
	(2881,1245)(2923,1258)(2967,1272)
	(3013,1287)(3061,1303)(3108,1318)
	(3155,1333)(3201,1348)(3246,1363)
	(3289,1377)(3330,1391)(3369,1404)
	(3406,1417)(3442,1429)(3475,1441)
	(3507,1453)(3538,1464)(3567,1475)
	(3596,1486)(3623,1497)(3650,1508)
	(3676,1519)(3702,1529)(3729,1541)
	(3757,1553)(3784,1565)(3811,1578)
	(3839,1591)(3867,1605)(3896,1619)
	(3926,1634)(3957,1650)(3989,1667)
	(4023,1685)(4058,1703)(4094,1723)
	(4132,1744)(4171,1765)(4210,1787)
	(4250,1809)(4289,1830)(4327,1852)
	(4362,1872)(4395,1891)(4425,1907)
	(4450,1922)(4471,1934)(4487,1943)
	(4499,1949)(4506,1954)(4510,1956)(4512,1957)
\put(552,1642){\makebox(0,0)[lb]{\(-1\)}}
\put(3597,1642){\makebox(0,0)[lb]{\(+1\)}}
\put(2577,67){\makebox(0,0)[lb]{\(l_3\)}}
\put(1992,2087){\makebox(0,0)[lb]{\(l_2\)}}
\put(3397,1097){\makebox(0,0)[lb]{\(l_1\)}}
\put(2262,1012){\makebox(0,0)[lb]{\(\xi_{\lambda}\)}}
\end{picture}
}
\caption{}\label{fig5}
\end{figure}

Let \(\lambda\) lie in a small \mbox{\(\tau\)-neigh}\-bour\-hood of the curve
\(\gamma_+\) outside some fixed neighbourhoods of the end points. Let
\(C_{\lambda}\) be the Stokes complex corresponding to the point \(\lambda\),
i.~e. \(\xi_{\lambda}\) is the knot-point of \(C_{\lambda}\)
(see~Figure~\ref{fig5}). One of the Stokes lines, say \(l_1\), passes near the
point \(1\) (\(l_1\) passes through \(1\) if \(\lambda\in\gamma_+\)). As
before, denote by \(\Omega\) the union of \(G\) and a
\mbox{\(\delta\)-neigh}\-bour\-hood of \([-1,1]\). Consider the canonical
domain \(\Omega_1\) lying in \(\Omega\) between the lines \(l_2\) and \(l_3\)
and containing the line \(l_1\), and the canonical domain \(\Omega_2\) in
\(\Omega\) lying between the lines \(l_3\) and \(l_1\) and containing \(l_2\).
Let \(v_j^+(z,\lambda)\), \(v_j^-(z,\lambda)\) be the pairs of solutions
having in canonical domains \(\Omega_j\), \(j=1,2\) asymptotics~\eqref{eq2:5}.
Asymptotics~\eqref{eq2:5} for the pair of solutions \(v_2^+\), \(v_2^-\) in the
domain \(\Omega_2\), does not remain valid in the domain \(\Omega_1\). However,
there is a connection formula for the solutions.
\begin{lem}\label{lem2:?}
The following transmission formula is valid: for \(z\in\Omega_1\) one has the
representation
\begin{equation}\label{eq2:4}
	\begin{pmatrix}v_2^+(z,\lambda)\\ v_2^-(z,\lambda)\end{pmatrix}=
	e^{i\pi/6}\begin{pmatrix}-i[1]&[1]\\1&0\end{pmatrix}
	\begin{pmatrix}v_1^+(z,\lambda)\\v_1^-(z,\lambda)\end{pmatrix},
\end{equation}
where \([1]=1+O(\varepsilon)\) and \(|O(\varepsilon)|<C\varepsilon\) with a
constant \(C\) depending on \(z\) and \(\lambda\). However, given any compact
\(K\) in \(\Omega_1=\Omega_1(\lambda)\) there exists a neighbourhood
\(U_{\lambda}\) of the point \(\lambda\) such that the last estimate holds for
all \(z\in K\) and all \(\lambda\) belonging to \(U_{\lambda}\) with a
constant \(C\) depending only on \(K\) and \(U_{\lambda}\).
\end{lem}
\begin{proof}
See the monographs~\cite{O} and~\cite{F}.
\end{proof}

Now, we can complete the proof of the theorem. Consider the characteristic
determinant
\[
	\Delta(\lambda)=\begin{vmatrix}
	v_2^+(-1,\lambda)& v_2^+(1,\lambda)\\v_2^-(-1,\lambda)&
	v_2^-(1,\lambda)\end{vmatrix}.
\]
For \(\lambda\) lying near the curve \(\gamma_+\) the points \(-1\) and \(+1\)
belong to the domains \(\Omega_2\) and \(\Omega_1\), respectively. Using
asymptotics~\eqref{eq2:5} and~\eqref{eq2:4} and abreviating by the term
\(e^{-i\pi/6}T(\lambda)\) (see the proof of Theorem~\ref{tm2:5}), we find
\[
	\Delta(\lambda)=\begin{vmatrix}
	[1]\exp(\varepsilon^{-1}S(-1,\lambda))&
	-i[1]\exp(\varepsilon^{-1}S(1,\lambda))+
	[1]\exp(-\varepsilon^{-1}S(1,\lambda))\\
	[1]\exp(-\varepsilon^{-1}S(-1,\lambda))&
	[1]\exp(\varepsilon^{-1}S(1,\lambda))\end{vmatrix}.
\]

The branch of the function \(S(z,\lambda)\) is defined by the condition \(\Re
S(-1,\lambda)=:\alpha(\lambda)<0\), if \(\lambda\) is located in the
\mbox{\(\tau\)-neigh}\-bour\-hood of \(\gamma_+\) outside \(U_{\delta}(b)\)
and \(U_{\delta}(\lambda_0)\). Notice, that \(\Re S(1,\lambda)\to 0\) as
\(\tau\to 0\). Hence, we can choose a number \(\tau>0\) such small that
\(\Re S(1,\lambda)<\alpha(\lambda)/2\). In this case the term
\(\exp\varepsilon^{-1}(S(-1,\lambda)+S(1,\lambda))\) decays exponentially,
while the term \(\exp(-\varepsilon^{-1}S(-1,\lambda))\) grows exponentially
as \(\varepsilon\to 0\).
So, the equation \(\Delta(\lambda)=0\) is equivalent up to exponentially small
terms to the equation
\[
	[1]e^{-\varepsilon^{-1}S(1,\lambda)}-
	i[1]e^{\varepsilon^{-1}S(1,\lambda)}=0,
\]
or
\begin{equation}\label{eq:cos}
	\cos\left(\dfrac{1}{i\varepsilon}\int\limits_{\xi_{\lambda}}^1
	\sqrt{i(q(\xi)-\lambda)}\,d\xi-\dfrac{\pi}{4}\right)=
	O(\varepsilon).
\end{equation}

Assume that the term \(O(\varepsilon)\) in this equation equals zero. Then,
the roots near the curve \(\gamma_+\) are determined explicitly by the
equations
\begin{equation}\label{eq2:9}
	-iQ^+(\lambda)=\varepsilon\left(k\pi-\pi/4\right),
	\quad k\in\mathbb Z.
\end{equation}
The function \(-iQ^+(\lambda)\) is real along the curve \(\gamma_+\) and
monotonous (see Lemma~\ref{lem2:3}). Hence, there exist integers \(p_+\) and
\(m_+\), such that for all \(p_+\leqslant k\leqslant m_+\)
equations~\eqref{eq2:9} have solutions \(\mu_k\) lying on \(\gamma_+\) outside
the \mbox{\(\delta\)-neigh}\-bour\-hoods of the end points \(b\) and
\(\lambda_0\). The existence of simple roots of perturbed
equation~\eqref{eq:cos} in \mbox{\(C\varepsilon^2\)-neigh}\-bour\-hoods of the
points \(\{\mu_k\}_{p_+}^{m_+}\) can be proved by standard means using the
Rouch\'e theorem (see details in Theorem~5.1 of~\cite{ShT}). The
representation for the counting eigenvalue function \(N(\lambda)\) along the
curve \(\gamma_+\) can be obtained from formulae~\eqref{eq2:9}. This can be
done in the same way as in Theorem~5.2 of~\cite{ShT}.

Certainly, the same assertions are valid for the eigenvalues near the curve
\(\gamma_-\). The proof of the formulae for the eigenvalues near the curve
\(\gamma_{\infty}\) can be realized simplier. Namely, there is no need to use
the transmission formulae to get asymptotic representation of the
characteristic determinant for all \(\lambda\) lying below both curved
\(\tilde\gamma_+\) and \(\tilde\gamma_-\). For such values of \(\lambda\) the
segment \([-1,1]\) lies between the left and the right Stokes lines, and we
can use asymptotics~\eqref{eq2:4} simultaneously in both points \(-1\) and
\(+1\). After simple calculations we obtain that for \(\lambda\) lying below
the curves \(\tilde\gamma_+\) and \(\tilde\gamma_-\) the equation
\(\Delta(\lambda)=0\) is equivalent to the following one
\[
	\sin\dfrac{1}{i\varepsilon}Q(\lambda)=O(\varepsilon).
\]
Using standard arguments we get the localization formulae for the eigenvalues
and the representation for the eigenvalue counting function along
\(\gamma_{\infty}\). Actually, a sharper analysis can be carried out. Namely,
the formulae
\begin{align*}
	\lambda_k&=\mu_k+\varepsilon^2\mu_k^{-1}O(1),&k&=s,s+1,\ldots,
\end{align*}
can be obtained for the eigenvalues near \(\gamma_{\infty}\) (here \(|O(1)|\)
is estimated by a constant \(C\) not dependent on \(k\) and \(\varepsilon\)).
To get these formulae, one has to use an analogue of Lemma~4.3 from
paper~\cite{ShT}. Here we omit details. Theorem is proved.
\end{proof}

\section{The case of Couette--Poiseuille profile}

\begin{figure}[t]
{
\unitlength=4cm
\begin{picture}(2.2,1.9)(-1.2,-1.8)

\scriptsize

\put(-1.1,0){\line(1,0){2.2}}
\put(0,0.1){\line(0,-1){1.7}}
\put(0.03,0.04){$0$}


\put(-1,0){\circle{0.03}}
\put(-1.03,0.04){$-1$}

\put(1,0){\circle{0.03}}
\put(1,0.04){$1$}

\put(0.333,-1.647){\circle*{0.015}}
\put(0.332,-1.590){\circle*{0.015}}
\put(0.332,-1.533){\circle*{0.015}}
\put(0.332,-1.478){\circle*{0.015}}
\put(0.332,-1.424){\circle*{0.015}}
\put(0.332,-1.370){\circle*{0.015}}
\put(0.332,-1.318){\circle*{0.015}}
\put(0.332,-1.267){\circle*{0.015}}
\put(0.332,-1.216){\circle*{0.015}}
\put(0.332,-1.167){\circle*{0.015}}
\put(0.332,-1.118){\circle*{0.015}}
\put(0.332,-1.070){\circle*{0.015}}
\put(0.331,-1.024){\circle*{0.015}}
\put(0.331,-0.978){\circle*{0.015}}
\put(0.815,-0.103){\circle*{0.015}}
\put(0.815,-0.103){\circle*{0.015}}
\put(0.331,-0.933){\circle*{0.015}}
\put(0.331,-0.889){\circle*{0.015}}
\put(0.331,-0.846){\circle*{0.015}}
\put(0.331,-0.804){\circle*{0.015}}
\put(0.681,-0.172){\circle*{0.015}}
\put(0.681,-0.172){\circle*{0.015}}
\put(0.330,-0.763){\circle*{0.015}}
\put(0.330,-0.722){\circle*{0.015}}
\put(0.330,-0.683){\circle*{0.015}}
\put(0.573,-0.223){\circle*{0.015}}
\put(0.573,-0.223){\circle*{0.015}}
\put(0.329,-0.645){\circle*{0.015}}
\put(0.329,-0.607){\circle*{0.015}}
\put(0.329,-0.570){\circle*{0.015}}
\put(0.480,-0.264){\circle*{0.015}}
\put(0.480,-0.264){\circle*{0.015}}
\put(0.328,-0.534){\circle*{0.015}}
\put(0.328,-0.500){\circle*{0.015}}
\put(0.327,-0.465){\circle*{0.015}}
\put(0.326,-0.432){\circle*{0.015}}
\put(0.396,-0.298){\circle*{0.015}}
\put(0.396,-0.298){\circle*{0.015}}
\put(0.326,-0.400){\circle*{0.015}}
\put(0.325,-0.368){\circle*{0.015}}
\put(0.327,-0.338){\circle*{0.015}}
\put(0.326,-0.321){\circle*{0.015}}
\put(0.312,-0.309){\circle*{0.015}}
\put(0.290,-0.290){\circle*{0.015}}
\put(0.270,-0.270){\circle*{0.015}}
\put(0.250,-0.250){\circle*{0.015}}
\put(0.230,-0.230){\circle*{0.015}}
\put(0.210,-0.210){\circle*{0.015}}
\put(0.190,-0.190){\circle*{0.015}}
\put(0.170,-0.170){\circle*{0.015}}
\put(0.150,-0.150){\circle*{0.015}}
\put(0.130,-0.130){\circle*{0.015}}
\put(0.110,-0.110){\circle*{0.015}}
\put(0.090,-0.090){\circle*{0.015}}
\put(0.070,-0.070){\circle*{0.015}}
\put(0.050,-0.050){\circle*{0.015}}
\put(0.030,-0.030){\circle*{0.015}}
\put(0.010,-0.010){\circle*{0.015}}
\end{picture}
}
\caption{}\label{fig6}
\end{figure}

Profiles of the form \(q(x)=\alpha x^2+\beta x+\gamma\),
\(\alpha,\beta,\gamma\in\mathbb R\), correspond to stationary solutions of the
Navier--Stokes equation, therefore, they have special interest in connection
with Orr--Sommerfeld problem~\eqref{eq1},~\eqref{eq2}. Consequently, the study
of model problem~\eqref{eq3},~\eqref{eq4} for functions \(q(x)\) of this form
is quite important. Since
\[
	\alpha x^2+\beta x+\gamma=\alpha\left(x+\dfrac{\beta}{2\alpha}
	\right)^2+\gamma-\dfrac{\beta^2}{4\alpha^2},
\]
the substitution of the spectral and small parameters
\[
	\varepsilon=\sqrt{\alpha}\varepsilon',\quad\lambda=\sqrt{\alpha}
	(\lambda'-\gamma+\beta^2/4\alpha^2)
\]
leads to model problem~\eqref{eq3},~\eqref{eq4} with the function
\(q(x)=(x-\beta')^2\), \(\beta'=\beta/2\alpha\). Further we write \(\beta\)
instead of \(\beta'\).

The cases \(\beta=0\) and \(\beta\neq 0\) are different and they have to be
treated separately. In the first case (i.~e. for \(q(x)=x^2\)) the
eigenfunctions are either even or odd (see~\cite{DR}, for example).
Consequently, the spectrum of problem~\eqref{eq3},~\eqref{eq4} with the
function \(q(x)=x^2\) consists of the spectra of two problems on the segment
\([0,1]\):
\begin{equation}\label{eq3:1}
\begin{array}{c@{\hspace{0.2\textwidth}}c}
	i\varepsilon^2 y''+x^2 y''=\lambda y&
	i\varepsilon^2 y''+x^2 y''=\lambda y\\
	y(0)=y(1)=0,& y'(0)=y(1)=0.
\end{array}
\end{equation}
The first of these problems is solved in~\S\,\ref{par1}; the second one can be
solved in the same way: the replacement of the boundary condition \(y(0)=0\)
by the condition \(y'(0)=0\) does not change the essence of the matter. The
method for the solution remains the same. However, the problem
\begin{equation}\label{eq3:2}
\begin{gathered}
	i\varepsilon y''+x^2 y=\lambda y,\\
	y'(-1)=y(1)=0
\end{gathered}
\end{equation}
with non-symmetrical boundary conditions can not be reduced to the study of
similar problems on the segment \([0,1]\) with the monotonous function
\(x^2\). The solution of~\eqref{eq3:2} is essentially more difficult. The
method for the study of this problem is proposed in~\cite{ShT} (formally, the
Dirichlet boundary conditions are treated in~\cite{ShT}, however, the problem
is studied globally on \([-1,1]\), and the proposed method can be used without
changes for arbitrary separated boundary conditions). It is easy to see
analyzing this method, that the form of the limit spectral graph \(\Gamma\)
for the function \(q(x)=x^2\) and the formulae for the counting eigenvalue
functions along the curves do not depend on boundary conditions. Hence, both
problems in~\eqref{eq3:1} and problem~\eqref{eq3:2} have the same limit
spectral graph (see~Figure~\ref{fig6}) and the same eigenvalue counting
functions up to multiplication by the coefficient
\(2\).

Actually, the method of~\cite{ShT} works to solve the problem
\begin{equation}\label{eq3:3}
\begin{gathered}
	i\varepsilon^2 y''+(x-\beta)^2 y=\lambda y,\\
	y(-1)=y(1)=0,
\end{gathered}
\end{equation}
with \(\beta\neq 0\). However, the limit spectral graph for this problem takes
a more complicated form. The same is true for the eigenvalue formulae and the
counting eigenvalue functions along the curves of the limit graph.

\begin{figure}[t]
{
\unitlength=4cm
\begin{picture}(2.2,1.9)(-1.2,-1.8)

\scriptsize

\put(-1.1,0){\line(1,0){2.2}}
\put(0,0.1){\line(0,-1){1.7}}
\put(0.03,0.04){$0$}


\put(-1,0){\circle{0.03}}
\put(-1.03,0.04){$-1$}

\put(0.5625,0){\circle{0.03}}
\put(0.6,0.04){$9/16$}

\put(1,0){\circle{0.03}}
\put(1,0.04){$1$}
\put(0.270,-1.650){\circle*{0.015}}
\put(0.270,-1.593){\circle*{0.015}}
\put(0.270,-1.537){\circle*{0.015}}
\put(0.270,-1.482){\circle*{0.015}}
\put(0.270,-1.427){\circle*{0.015}}
\put(0.270,-1.374){\circle*{0.015}}
\put(0.270,-1.322){\circle*{0.015}}
\put(0.270,-1.271){\circle*{0.015}}
\put(0.269,-1.220){\circle*{0.015}}
\put(0.269,-1.171){\circle*{0.015}}
\put(0.269,-1.122){\circle*{0.015}}
\put(0.269,-1.075){\circle*{0.015}}
\put(0.831,-0.095){\circle*{0.015}}
\put(0.269,-1.029){\circle*{0.015}}
\put(0.269,-0.983){\circle*{0.015}}
\put(0.269,-0.938){\circle*{0.015}}
\put(0.707,-0.159){\circle*{0.015}}
\put(0.268,-0.895){\circle*{0.015}}
\put(0.268,-0.852){\circle*{0.015}}
\put(0.268,-0.810){\circle*{0.015}}
\put(0.608,-0.207){\circle*{0.015}}
\put(0.268,-0.769){\circle*{0.015}}
\put(0.268,-0.729){\circle*{0.015}}
\put(0.267,-0.690){\circle*{0.015}}
\put(0.523,-0.246){\circle*{0.015}}
\put(0.267,-0.652){\circle*{0.015}}
\put(0.267,-0.615){\circle*{0.015}}
\put(0.445,-0.278){\circle*{0.015}}
\put(0.266,-0.579){\circle*{0.015}}
\put(0.266,-0.544){\circle*{0.015}}
\put(0.265,-0.509){\circle*{0.015}}
\put(0.374,-0.306){\circle*{0.015}}
\put(0.265,-0.475){\circle*{0.015}}
\put(0.424,-0.076){\circle*{0.015}}
\put(0.264,-0.443){\circle*{0.015}}
\put(0.263,-0.411){\circle*{0.015}}
\put(0.307,-0.329){\circle*{0.015}}
\put(0.263,-0.380){\circle*{0.015}}
\put(0.262,-0.352){\circle*{0.015}}
\put(0.253,-0.328){\circle*{0.015}}
\put(0.324,-0.125){\circle*{0.015}}
\put(0.240,-0.302){\circle*{0.015}}
\put(0.227,-0.276){\circle*{0.015}}
\put(0.214,-0.251){\circle*{0.015}}
\put(0.245,-0.159){\circle*{0.015}}
\put(0.202,-0.225){\circle*{0.015}}
\put(0.191,-0.200){\circle*{0.015}}
\put(0.182,-0.180){\circle*{0.015}}
\put(0.167,-0.166){\circle*{0.015}}
\put(0.149,-0.149){\circle*{0.015}}
\put(0.131,-0.131){\circle*{0.015}}
\put(0.114,-0.114){\circle*{0.015}}
\put(0.096,-0.096){\circle*{0.015}}
\put(0.079,-0.079){\circle*{0.015}}
\put(0.061,-0.061){\circle*{0.015}}
\put(0.044,-0.044){\circle*{0.015}}
\put(0.026,-0.026){\circle*{0.015}}
\put(0.009,-0.009){\circle*{0.015}}
\end{picture}
}
\caption{}\label{fig7}
\end{figure}

Let us formulate the main results concerning problem~\eqref{eq3:3}. Without
loss of generality we assume, that \(\beta\in(0,1)\). Set
\[
	a=(-1-\beta)^2,\quad b=(1-\beta)^2.
\]
Repeating simple arguments of Lemma~\ref{lem1:1}, we obtain, that the spectrum
of problem~\eqref{eq3:3} lies in the semistrip
\[
	\Pi=\left\{\lambda\;\vline\;\Im\lambda<0,\;0<\Re\lambda<b\right\}.
\]
Consider the following curves in \(\Pi\):
\begin{align*}
	\tilde\gamma_0&=\left\{\lambda\in\Pi\;\vline\;\arg\lambda=-\pi/4
	\right\},\\
	\tilde\gamma_b&=\left\{\lambda\in\Pi\;\vline\;
	\Re\int\limits_{\sqrt{\lambda}+\beta}^1\sqrt{i(q(\xi)-\lambda)}\,
	d\xi=0\right\},\\
	\tilde\gamma_a&=\left\{\lambda\in\Pi\;\vline\;
	\Re\int\limits_{-\sqrt{\lambda}+\beta}^{-1}\sqrt{i(q(\xi)-\lambda)}
	\,d\xi=0\right\},\\
	\tilde\gamma_-&=\left\{\lambda\in\Pi\;\vline\;
	\Re\int\limits_{\sqrt{\lambda}+\beta}^{-1}\sqrt{i(q(\xi)-\lambda)}
	\,d\xi=0\right\},\\
	\tilde\gamma_{\infty}&=\left\{\lambda\in\Pi\;\vline\;
	\Re\int\limits_{-1}^1\sqrt{i(q(\xi)-\lambda)}\,
	d\xi=0\right\},
\end{align*}
with \(q(\xi)=(\xi-\beta)^2\). It is easily seen that all these curves have no
self-intersections (see Lemma~\ref{lem2:6}).
\begin{lem}\label{lem3:1}
The curves \(\tilde\gamma_0\), \(\tilde\gamma_b\) and \(\tilde\gamma_-\) have
the only intersection point \(\lambda_1\in\Pi\). The curves
\(\tilde\gamma_-\), \(\tilde\gamma_a\) and \(\tilde\gamma_{\infty}\) intersect
in a point \(\lambda_2\in\Pi\). There are no other intersection points of all
these curves.
\end{lem}
\begin{proof}
It can be carried out using the ideas from Lemma~\ref{lem2:6}.
\end{proof}

Denote by \(\gamma_0\) the part of \(\tilde\gamma_0\) between 0 and
\(\lambda_1\) (i.~e. \(\gamma_0=[0,\lambda_1]\)), by \(\gamma_b\) the part of
\(\tilde\gamma_b\) between the points \(b\) and \(\lambda_1\), by \(\gamma_-\)
the part of \(\tilde\gamma_-\) between the points \(\lambda_2\) and \(a\), and
by \(\gamma_{\infty}\) the part of \(\tilde\gamma_{\infty}\) between
\(\lambda_2\) and \(-i\infty\).
\begin{tm}\label{tm3:1}
Given small \(\tau>0\) there is a number \(\varepsilon_0=\varepsilon_0(\tau)\)
such that for all \(\varepsilon<\varepsilon_0\) the
eigenvalues of problem~\eqref{eq3:3} lie inside the
\mbox{\(\tau\)-neigh}\-bour\-hood of the set
\begin{equation}\label{eq:Ga}
	\Gamma=\gamma_0\cup\gamma_a\cup\gamma_-\cup\gamma_b\cup
	\gamma_{\infty}.
\end{equation}
\end{tm}
\begin{proof}
The first step in proving of this theorem is to clarify the geometry of the
Stokes graphs of the Weber equation~\eqref{eq3:3}. Actually, this work was
carried out in~\cite{ShT}. There are tree reasons for points \(\lambda\in\Pi\)
to be accumulation points for the eigenvalues as \(\varepsilon\to 0\). First,
at least one point \(+1\) or \(-1\) belongs to the Stokes graph
\(C_{\lambda}\) of equation~\eqref{eq3:3}. The set of such points form the
lines which we call \textit{singular}. In our case the lines
\(\tilde\gamma_a\), \(\tilde\gamma_-\) and \(\tilde\gamma_b\) are singular.
However, not all points of a singular line \(\tilde\gamma\) belong to the
limit spectral graph. We have to exclude the points \(\mu\in\tilde\gamma\)
possessing the following property: there is a path in the \mbox{\(z\)-pla}\-ne
connecting the points \(-1\) and \(+1\) such that it intersect only one Stokes
line of all Stokes complexes \(C_{\lambda}\) for all \(\lambda\in
U_{\delta}(\mu)\) with sufficiently small \(\delta>0\). In our case after
elimination of such points we get the curves \(\gamma_a\), \(\gamma_-\) and
\(\gamma_b\). Second, there are points \(\mu\) in the
\mbox{\(\lambda\)-pla}\-ne such that the geometry of the Stokes graphs is not
preserved in any sufficiently small neighbourhood \(U_{\delta}(\mu)\) (see
details in~\cite{ShT}). The points of such kind form lines which we call
\textit{critical}. In our case there is the only critical line
\(\tilde\gamma_0\) (for \(\mu\in\tilde\gamma_0\) the Stokes graph consists of
one complex, while for \(\mu\not\in\tilde\gamma_0\) it consists of two
complexes). Again, we have to exclude the points of critical lines which
possess the property that we described before. Then, in our case we get the
curve \(\gamma_0\). Finally, the curve
\[
	\tilde\gamma_{\infty}=\left\{\lambda\;\vline\;\Re\int\limits_{-1}^1
	\sqrt{i(q(x)-\lambda)}\,dx=0\right\}
\]
we call \textit{the main} line. The part of this line between \(-i\infty\) and
the first intersection point with singular or critical lines has to be
included in the limit spectral graph.

This introduction gives an understanding, how to prove Theorem~\ref{tm3:1}.
Here a complete proof is omitted. The details can be found in~\cite{ShT3}.
\end{proof}
\begin{tm}
The set~\eqref{eq:Ga} is the limit spectral graph of problem~\eqref{eq3:3},
i.~e. the points \(\lambda\in\Gamma\) and only these points are the
accumulation points of the eigenvalues as \(\varepsilon\to 0\). Given
\(\delta>0\) there are numbers \(\varepsilon_0=\varepsilon_0(\delta)\) and
\(C=C(\delta)\) such that for all
\(\varepsilon<\varepsilon_0\) the eigenvalues of
problem~\eqref{eq3:3} lie in
the set
\[
	U_{\delta}=U_{\delta}(0)\cup U_{\delta}(a)\cup U_{\delta}(b)\cup
	U_{\delta}(\lambda_1)\cup U_{\delta}(\lambda_2)
\]
and in the \mbox{\(C\varepsilon^2\)-neigh}\-bour\-hoods of the points
\(\mu_k\in\Gamma\) which are determined by the equations
\[
\begin{array}{c@{\hspace{2em}}l@{\hspace{0.7em}}l}
	i\int\limits_{\sqrt{\lambda}+\beta}^1\sqrt{i(q(\xi)-\lambda)}\,d\xi
	=\varepsilon(\pi k-\pi/4),& \mu_k\in\gamma_b,&
	k\in\mathbb Z,\\
	i\int\limits_{-\sqrt{\lambda}+\beta}^{-1}\sqrt{i(q(\xi)-\lambda)}
	\,d\xi=\varepsilon(\pi k-\pi/4),& \mu_k\in\gamma_a,&
	k\in\mathbb Z,\\
	i\int\limits_{\sqrt{\lambda}+\beta}^{-1}\sqrt{i(q(\xi)-\lambda)}
	\,d\xi=\varepsilon(\pi k-\pi/4),& \mu_k\in\gamma_-,&
	k\in\mathbb Z,\\
	i\int\limits_{-1}^1\sqrt{i(q(\xi)-\lambda)}
	\,d\xi=\varepsilon\pi k,& \mu_k\in\gamma_{\infty},&
	k\in\mathbb Z,\\
	\lambda_k^0=(2k+1)\varepsilon e^{-i\pi/4},&\mu_k
	\in\gamma_0,&k\in\mathbb Z.
\end{array}
\]
The \mbox{\(C\varepsilon^2\)-neigh}\-bour\-hoods of all points
\(\mu_k\in\Gamma\setminus U_{\delta}\) contain only one simple eigenvalue. The
counting eigenvalue functions along the curves of the graph \(\Gamma\) have
representations
\begin{align*}
	N(\lambda)&=\dfrac{1}{i\pi\varepsilon}\int\limits_{\sqrt{\lambda}+
	\beta}^1\sqrt{i(q(\xi)-\lambda)}\,d\xi+O(1),&\lambda&\in\gamma_b,\\
	N(\lambda)&=\dfrac{1}{i\pi\varepsilon}\int\limits_{-\sqrt{\lambda}+
	\beta}^{-1}\sqrt{i(q(\xi)-\lambda)}\,d\xi+O(1),&\lambda&\in\gamma_a,\\
	N(\lambda)&=\dfrac{1}{i\pi\varepsilon}\int\limits_{\sqrt{\lambda}+
	\beta}^{-1}\sqrt{i(q(\xi)-\lambda)}\,d\xi+O(1),&\lambda&\in\gamma_-,\\
	N(\lambda)&=\dfrac{1}{i\pi\varepsilon}\int\limits_{-1}^1
	\sqrt{i(q(\xi)-\lambda)}\,d\xi+O(1),&\lambda&\in\gamma_{\infty},\\
	N(\lambda)&=\dfrac{1}{2\varepsilon}e^{i\pi/4}\lambda,&
	\lambda&\in\gamma_0.
\end{align*}
\end{tm}
\begin{proof}
The details of the proof of this theorem can be found in~\cite{ShT3}. The
spectral portrait of problem~\eqref{eq3:3} for the function
\(q(x)=(7x-1)^2/64\), \(\varepsilon^2=5000\) is
depicted in Figure~\ref{fig7}.
\end{proof}

\begin{figure}[t]
{
\unitlength=4cm
\begin{picture}(2.2,1.9)(-1.2,-1.8)

\scriptsize

\put(-1.1,0){\line(1,0){2.2}}
\put(0,0.1){\line(0,-1){1.7}}
\put(0.03,0.04){$0$}


\put(-1,0){\circle{0.03}}
\put(-1.03,0.04){$-1$}

\put(1,0){\circle{0.03}}
\put(1,0.04){$1$}

\put(0.000,-1.542){\circle*{0.015}}
\put(0.783,-0.207){\circle*{0.015}}
\put(0.000,-1.468){\circle*{0.015}}
\put(0.722,-0.080){\circle*{0.015}}
\put(-0.000,-1.396){\circle*{0.015}}
\put(0.616,-0.292){\circle*{0.015}}
\put(0.000,-1.325){\circle*{0.015}}
\put(0.567,-0.181){\circle*{0.015}}
\put(-0.000,-1.255){\circle*{0.015}}
\put(0.000,-1.187){\circle*{0.015}}
\put(0.477,-0.365){\circle*{0.015}}
\put(-0.000,-1.121){\circle*{0.015}}
\put(0.438,-0.263){\circle*{0.015}}
\put(0.000,-1.056){\circle*{0.015}}
\put(0.355,-0.430){\circle*{0.015}}
\put(-0.000,-0.992){\circle*{0.015}}
\put(0.322,-0.335){\circle*{0.015}}
\put(0.000,-0.930){\circle*{0.015}}
\put(0.243,-0.490){\circle*{0.015}}
\put(-0.000,-0.869){\circle*{0.015}}
\put(0.215,-0.401){\circle*{0.015}}
\put(0.000,-0.810){\circle*{0.015}}
\put(0.139,-0.547){\circle*{0.015}}
\put(-0.001,-0.752){\circle*{0.015}}
\put(0.003,-0.698){\circle*{0.015}}
\put(0.114,-0.460){\circle*{0.015}}
\put(0.043,-0.602){\circle*{0.015}}
\put(-0.003,-0.639){\circle*{0.015}}
\put(0.020,-0.525){\circle*{0.015}}
\put(-0.042,-0.601){\circle*{0.015}}
\put(-0.019,-0.524){\circle*{0.015}}
\put(-0.139,-0.547){\circle*{0.015}}
\put(-0.115,-0.462){\circle*{0.015}}
\put(-0.243,-0.490){\circle*{0.015}}
\put(-0.215,-0.401){\circle*{0.015}}
\put(-0.355,-0.430){\circle*{0.015}}
\put(-0.322,-0.335){\circle*{0.015}}
\put(-0.477,-0.365){\circle*{0.015}}
\put(-0.438,-0.263){\circle*{0.015}}
\put(-0.567,-0.181){\circle*{0.015}}
\put(-0.616,-0.292){\circle*{0.015}}
\put(-0.722,-0.080){\circle*{0.015}}
\put(-0.783,-0.207){\circle*{0.015}}
\end{picture}
}
\caption{}\label{fig8}
\end{figure}

\section{The Orr--Sommerfeld problem}
The description of global behaviour of the spectrum of the Orr--Sommerfeld
problem as \(R\to\infty\) have been carried out only for the functions
\(q(x)=x\) (Couette profile) and \(q(x)=x^2\) or \(q(x)=1-x^2\) (Poiseuille
profile). Here we point out the papers~\cite{Ch},~\cite{DSh2}, \cite{M}
and~\cite{ShT2}. Actually, the methods developed in~\cite{ShT2}
and~\cite{ShT3} can be used without essential changes to treat the
Couette--Poiseuille profiles \(q(x)=\alpha x^2+\beta x+\gamma\). Here we shall
only formulate the result obtained in~\cite{DSh2} and the generalization of
the result~\cite{ShT2}; the detailed proof requires a serious work.
Figure~\ref{fig8} (it is borrowed from the paper~\cite{ShNz}) gives an
illustration of the first theorem of this section. Here the spectrum is
calculated for \(\alpha=1\) and \(R=4000\).

It was shown in Introduction, that in the case of the Couette profile
\(q(x)=x\) the Orr--Sommerfeld problem takes the form
\begin{equation}\label{eq:4.1}
	\begin{gathered}
		-i\varepsilon z''+(i\varepsilon\alpha^2-x)z=
		\lambda z,\qquad\varepsilon=1/\alpha R,\\
		\int\limits_{-1}^1\sinh[\alpha(1-t)]\,z(t)\,dt=
		\int\limits_{-1}^1\cosh[\alpha(1-t)]\,z(t)\,dt=0.
	\end{gathered}
\end{equation}
To describe the eigenvalue behaviour of this problem as \(\varepsilon\to 0\),
we consider the rectangular coordinate system  \(\{t,\gamma\}\) in the
\mbox{\(\lambda\)-comp}\-lex plane taking the point \(-1\) as the origin and
directing the axis \(t\) along the segment \([-1,-i/\sqrt3]\)
(see~Figure~\ref{fig10}).

Set
\begin{align*}
	c(t)&=2\sqrt{\pi}\dfrac{\left|\sinh\left(\alpha\left(2-e^{-i\pi/4}t
	\right)\right)\right|}{\sinh 2\alpha},&\varphi(t)&=
	\dfrac{1}{2\pi}\arg\sinh\left(\alpha\left(2-e^{-i\pi/4}t\right)
	\right),
\end{align*}
where the main branch of the argument is chosen.

\begin{figure}[t]
\begin{center}
\setlength{\unitlength}{0.00035333in}
\begingroup\makeatletter\ifx\SetFigFont\undefined%
\gdef\SetFigFont#1#2#3#4#5{%
  \reset@font\fontsize{#1}{#2pt}%
  \fontfamily{#3}\fontseries{#4}\fontshape{#5}%
  \selectfont}%
\fi\endgroup%
{\renewcommand{\dashlinestretch}{30}
\begin{picture}(8424,7989)(0,-10)
\put(312,4812){\circle*{150}}
\put(8112,4812){\circle*{150}}
\drawline(12,4812)(8412,4812)
\drawline(312,4812)(6162,1137)
\blacken\drawline(6044.428,1175.431)(6162.000,1137.000)(6076.345,1226.237)(6044.428,1175.431)
\drawline(312,4812)(2562,7962)
\blacken\drawline(2516.663,7846.915)(2562.000,7962.000)(2467.839,7881.789)(2516.663,7846.915)
\drawline(4212,7887)(4212,12)
\drawline(8112,4812)(4212,2337)
\dashline{150.000}(1662,4812)
	(2532,3912)(3541,3051)
	(4151,2605)(4212,2562)
\dashline{150.000}(387,3687)
	(1049,3551)(2607,2961)
	(3413,2581)(4212,2187)
\dashline{150.000}(1887,5112)
	(1809,4563)(1248,3691)
	(682,3420)(387,3312)
\dashline{75.000}(4212,2712)(4062,2712)(3912,2562)
	(3837,2487)(3837,2262)(3912,2112)
	(4062,1962)(4212,1962)(4362,1962)
	(4512,2112)(4587,2262)(4587,2487)
	(4437,2637)(4287,2712)
\put(8187,5112){\makebox(0,0)[lb]{\(1\)}}
\put(17,5112){\makebox(0,0)[lb]{\(-1\)}}
\put(1962,7662){\makebox(0,0)[lb]{\(\gamma\)}}
\put(5587,1087){\makebox(0,0)[lb]{\(t\)}}
\end{picture}
}
\end{center}
\caption{}\label{fig10}
\end{figure}

Consider in the \(\{t,\gamma\}\) plane the curves
\[
	\gamma_{\pm}(t)=\pm\dfrac{\varepsilon^{1/2}}{t^{1/2}}
	\ln\dfrac{c(t)t^{3/4}}{\varepsilon^{1/4}},\qquad t>0,
\]
and fix at these curves the points
\[
	\mu_k^{\pm}=\{t_k^{\pm},\gamma_{\pm}(t_k)\},
\]
where
\begin{align*}
	t_k^{\pm}&=\varepsilon^{1/3}\left(
	3\pi\left[k-1/4\mp\varphi((3\pi
	\varepsilon^{1/2}k)^{2/3})\right]\right)^{2/3},&
	k_0&\leqslant k\leqslant k_1,
\end{align*}
and the indices \(k_0,k_1\) are chosen in such a way that
\begin{align*}
	\varepsilon^{1/3}|\ln\varepsilon|&\leqslant t_k\leqslant2/\sqrt3-
	\dfrac23\left(\dfrac34\right)^{3/4}\varepsilon^{1/2}
	|\ln\varepsilon|,&k_0&\leqslant k\leqslant k_1.
\end{align*}

\begin{tm}\label{tm4:1}
Denote by \(U_k^{\pm}\) the neighbourhoods of the points \(\mu_k^{\pm}\) of
radius
\(\delta_k=C\varepsilon^{3/4}t_k^{-5/4}\)
and by \(\hat U_k^{\pm}\) the
symmetrical reflections of \(U_k^{\pm}\) with respect to the imaginary axis.
Denote by \(U_{\pm 1}\) the
\mbox{\(\varepsilon^{1/3}|\ln\varepsilon|\)-neigh}\-bour\-hoods of the points
\(\pm 1\) and by \(U_0\) the \mbox{\(2/3\cdot(3/4)^{3/4}\varepsilon^{1/2}
|\ln\varepsilon|\)-neigh}\-bour\-hood of the knot-point \(-i/\sqrt3\). Then,
there are numbers \(C>0\) and \(\varepsilon_0>0\)
such that all the eigenvalues of
problem~\eqref{eq:4.1} located near the segments \([\pm 1,-i/\sqrt3]\) lie
inside the circles \(U_{\pm 1}\), \(U_0\) and
\(\{U_k^{\pm}\}_{k_0-1}^{k_1+1}\), \(\{\hat U_k^{\pm}\}_{k_0-1}^{k_1+1}\),
provided that \(\varepsilon\leqslant\varepsilon_0\). The circles \(U_k^{\pm}\)
and \(\hat U_k^{\pm}\) contain only one simply eigenvalue. All the other
eigenvalues lie at the imaginary axis below the point \(-i/\sqrt3\) and have
representation
\begin{align*}
	\lambda_k&=-i(\rho_k+\varepsilon O(1)),&k=k_0,k_0+1,\ldots,
\end{align*}
where the numbers \(\rho_k\) and \(k_0\) are the same as in
Theorem~\ref{tm1:1}.
\end{tm}
\begin{proof}
As in Theorem~\ref{tm1:1}, special properties of the Airy functions are
explored in the proof. Details can be found in~\cite{DSh2}.
\end{proof}

\begin{figure}[t]
{
\unitlength=4cm
\begin{picture}(2.2,1.9)(-1.2,-1.8)

\scriptsize

\put(-1.1,0){\line(1,0){2.2}}
\put(0,0.1){\line(0,-1){1.7}}
\put(0.03,0.04){$0$}


\put(-1,0){\circle{0.03}}
\put(-1.03,0.04){$-1$}

\put(1,0){\circle{0.03}}
\put(1,0.04){$1$}

\put(0.330,-1.591){\circle*{0.015}}
\put(0.330,-1.534){\circle*{0.015}}
\put(0.330,-1.479){\circle*{0.015}}
\put(0.330,-1.425){\circle*{0.015}}
\put(0.330,-1.371){\circle*{0.015}}
\put(0.330,-1.319){\circle*{0.015}}
\put(0.330,-1.267){\circle*{0.015}}
\put(0.331,-1.216){\circle*{0.015}}
\put(0.331,-1.167){\circle*{0.015}}
\put(0.331,-1.118){\circle*{0.015}}
\put(0.331,-1.070){\circle*{0.015}}
\put(0.331,-1.024){\circle*{0.015}}
\put(0.331,-0.978){\circle*{0.015}}
\put(0.760,-0.215){\circle*{0.015}}
\put(0.331,-0.933){\circle*{0.015}}
\put(0.731,-0.243){\circle*{0.015}}
\put(0.330,-0.889){\circle*{0.015}}
\put(0.330,-0.846){\circle*{0.015}}
\put(0.732,-0.002){\circle*{0.015}}
\put(0.330,-0.804){\circle*{0.015}}
\put(0.330,-0.763){\circle*{0.015}}
\put(0.664,-0.060){\circle*{0.015}}
\put(0.330,-0.722){\circle*{0.015}}
\put(0.330,-0.683){\circle*{0.015}}
\put(0.580,-0.163){\circle*{0.015}}
\put(0.531,-0.279){\circle*{0.015}}
\put(0.330,-0.644){\circle*{0.015}}
\put(0.515,-0.317){\circle*{0.015}}
\put(0.329,-0.607){\circle*{0.015}}
\put(0.329,-0.570){\circle*{0.015}}
\put(0.328,-0.534){\circle*{0.015}}
\put(0.497,-0.169){\circle*{0.015}}
\put(0.328,-0.499){\circle*{0.015}}
\put(0.327,-0.465){\circle*{0.015}}
\put(0.422,-0.257){\circle*{0.015}}
\put(0.327,-0.431){\circle*{0.015}}
\put(0.326,-0.399){\circle*{0.015}}
\put(0.345,-0.372){\circle*{0.015}}
\put(0.352,-0.333){\circle*{0.015}}
\put(0.321,-0.328){\circle*{0.015}}
\put(0.316,-0.347){\circle*{0.015}}
\put(0.354,-0.249){\circle*{0.015}}
\put(0.290,-0.285){\circle*{0.015}}
\put(0.285,-0.291){\circle*{0.015}}
\put(0.250,-0.247){\circle*{0.015}}
\put(0.250,-0.247){\circle*{0.015}}
\put(0.210,-0.208){\circle*{0.015}}
\put(0.211,-0.207){\circle*{0.015}}
\put(0.170,-0.168){\circle*{0.015}}
\put(0.171,-0.168){\circle*{0.015}}
\put(0.130,-0.129){\circle*{0.015}}
\put(0.130,-0.129){\circle*{0.015}}
\put(0.090,-0.089){\circle*{0.015}}
\put(0.090,-0.089){\circle*{0.015}}
\put(0.050,-0.050){\circle*{0.015}}
\put(0.050,-0.050){\circle*{0.015}}
\end{picture}
}
\caption{}\label{fig9}
\end{figure}

\begin{tm}\label{tm4:2}
Given \(\tau>0\) there is \(\varepsilon=\varepsilon_0(\tau)\), such that
for \(\varepsilon<\varepsilon_0\) all the eigenvalues of
the Orr--Sommerfeld
problem~\eqref{eq1},~\eqref{eq2} with the Couette--Poiseuille profile
\(q(x)=(x-\beta)^2\), \(\beta\in(-1,1)\), lie in the
\mbox{\(\tau\)-neigh}\-bour\-hood of the limit spectral graph \(\Gamma\) of the
corresponding model problem. The main terms of the counting eigenvalue
functions along the curves of the graph \(\Gamma\) have the representations
given in Theorem~\ref{tm3:1}.
\end{tm}
\begin{proof} In the case \(\beta=0\) the proof is obtained in~\cite{ShT3}. For
\(\beta\neq 0\) the proof remains essentially the same, provided that the
analysis of the model problem with \(q(x)=(x-\beta)^2\) is
carried out
(see~\cite{ShT3}). The proof of the results about the counting eigenvalue
functions uses the tauberian technique developed in~\cite{Sh2}.
Figure~\ref{fig9} shows the spectrum of the Orr--Sommerfeld problem for
\(q(x)=x^2\), \(\alpha=1\) and \(R=3000\).
\end{proof}

\textbf{Acknowledgements.} The  author thanks prof.~J.~M.~Ball~(Oxford),
E.~B.~Davies~(Kings College London), D.~G.~Vassiliev~(Bath),
W.~D.~Evans~(Cardiff) for the interest to the problem
and for the invitation in November of~2001 to give talks
about spectral portraits of non-selfadjoint problems at their seminars. The
author is gratefull to prof.~M.~Brown~(Cardiff), S.~J.~Chapman~(Oxford),
E.~B.~Davies~(Kings College London), L.~Greenberg
(Maryland), M.~Marletta~(Cardiff),
L.~N.~Trefethen~(Oxford) for fruitfull discussions on a model and the
Orr--Sommerfeld operators and for the acquaintance with their works. The
author kindly thanks his russian collegues
prof.~D.~G.~Georgievskii, S.~Yu.~Dobrokhotov,
L.~A.~Kalyadin, A.~M.~Il'in, V.~P.~Maslov,
S.~N.~Naboko, N.~N.~Nefedov, D.~P.~Popov and
V.~I.~Zhuk for the interest to this subject and useful discussions.

\end{document}